\documentclass[journal]{IEEEtran}
\usepackage[numbers,sort&compress]{natbib}
\usepackage{amsfonts}
\usepackage[cmex10]{amsmath}
\usepackage{array}
\usepackage{mdwmath}
\usepackage{mdwtab}
\usepackage{eqparbox}
\usepackage[font=small]{caption}
\usepackage{stfloats}
\usepackage{url}
\usepackage{amssymb}
\usepackage{float}
\usepackage{indentfirst}
\usepackage{makeidx}
\usepackage{tabularx}
\usepackage{cases}
\usepackage{multirow}
\usepackage{booktabs}
\usepackage{diagbox}
\usepackage{mathtools}
\usepackage{color}
\usepackage{stfloats}
\usepackage{lipsum}
\usepackage{amsfonts}
\usepackage{bm}
\usepackage{dsfont}
\usepackage{lipsum}
\usepackage{amsthm}
\usepackage{graphicx, epsfig, subfigure}
\allowdisplaybreaks[4]

\makeatletter

\newif\if@restonecol
\makeatother

\usepackage[linesnumbered,ruled,vlined]{algorithm2e}
	\usepackage{algpseudocode}
	\usepackage{amsmath, bm}

\begin{document}
		\title{Distributed Learning over Networks with Graph-Attention-Based Personalization}
		\author{Zhuojun~Tian,
			Zhaoyang~Zhang,
			Zhaohui~Yang,
			Richeng~Jin, and Huaiyu~Dai
			\thanks{
				{The conference version of this paper has been accepted by IEEE ICC'23 Workshop on Edge Learning over 5G Mobile Networks and Beyond \cite{Tian2023}.}
				Z.~Tian (email: dankotian@zju.edu.cn), Z.~Zhang (Corresponding Author, email: ning\_ming@zju.edu.cn), Z.~Yang (email: yang\_zhaohui@zju.edu.cn) and R.~Jin (email: richengjin@zju.edu.cn) are with the College of Information Science and Electronic Engineering, Zhejiang University, Hangzhou, China, and also with Zhejiang Provincial Key Laboratory of Info. Proc., Commun. \& Netw. (IPCAN), Hangzhou 310027, China.}
			\thanks{H.~Dai (e-mail: huaiyu\_dai@ncsu.edu) is with the Department of Electrical and Computer Engineering, NC State University, USA.}
		}
		
		
		\maketitle
		
		\begin{abstract}
			In conventional distributed learning over a network, multiple agents collaboratively build a common machine learning model. 
			However, due to the underlying non-i.i.d. data distribution among agents, the unified learning model becomes inefficient for each agent to process its locally accessible data.
			To address this problem, we propose a graph-attention-based personalized training algorithm (GATTA) for distributed deep learning. The GATTA enables each agent to train its local personalized model while exploiting its correlation with neighboring nodes and utilizing their useful information for aggregation. 
			In particular, the personalized model in each agent is composed of a global part and a node-specific part.
			By treating each agent as one node in a graph and the node-specific parameters as its features, the benefits of the graph attention mechanism can be inherited.
			Namely, instead of aggregation based on averaging, it learns the specific weights for different neighboring nodes without requiring prior knowledge about the graph structure or the neighboring nodes' data distribution.
			Furthermore, relying on the weight-learning procedure, we develop a communication-efficient GATTA by skipping the transmission of information with small aggregation weights.
			Additionally, we theoretically analyze the convergence properties of GATTA for non-convex loss functions.
			Numerical results validate the excellent performances of the proposed algorithms in terms of convergence and communication cost.
		\end{abstract}

		\section{Introduction} \label{sec1}
		With the rapid development of deep learning as well as the growing storage and computational capacity of devices, distributed deep learning has attracted great attention recently. It can be widely applied in many areas such as cooperative localization in 5G networks, distributed signal processing and recommender system. In conventional distributed learning procedures, each agent has access to its own training data and cooperates with others to obtain a common global model. However, in practical scenarios, the agents distributed in different geographical locations always have their local partial view and tend to access data with heterogeneous distributions, i.e., the data distribution is non-i.i.d. for different agents. 
		Take the collaborative location problem as an example. The base stations located in different positions may have diverse surroundings, leading to different data distributions and projections from the input such as {channel state information (CSI)} to the output user location.
		In such non-i.i.d. conditions, the consensus model shared among all agents may have poor performance for the locally accessible data in each agent. {This problem motivates us to address the challenge of statistical heterogeneity in distributed learning, through developing the personalized model for each agent.} 
		
		In this work, we investigate the decentralized communication network, which does not require a central server and is thus more robust by removing the heavy communication burden concentrated on the central server. In every round of the decentralized learning, each agent executes a local update of the model and then shares the updated model with neighboring nodes for aggregation. A dedicated aggregation procedure is expected to utilize the effective information from neighboring nodes, which however is always implicit and difficult to be explicitly characterized in the non-i.i.d. scenario. Besides, the well-known decentralized stochastic gradient descent (D-SGD) \cite{Lian2017} aggregates the model parameters through averaging or weighted averaging, which can lead to performance loss since this aggregation does not take account of the non-i.i.d. data distribution.
		Thus, it is necessary and appealing to conceive an aggregation procedure for non-i.i.d. conditions in distributed learning over a network.
		Moreover, the aggregation procedure can be utilized to further reduce the communication cost during the training process.
		
		Recently, we have witnessed significant progress in solving non-i.i.d. challenges of Federated Learning (FL) \cite{Smith2017, Jiang2019, Mikhail2019, wang2019,  Kulkarni2020, yu2020, Fallah2020, Hanzely2020, Deng2020, Mansour2020, Ghosh2020, Dinh2020, Li2021, Caldarola2021, Collins2021, Niko2022, Fedbn2021, Felix2020}. FL is a centralized learning framework \cite{FL1, FL3} requiring a central server for model aggregation.
		The experiments and analysis in \cite{Zhao2018} show the significant performance degradation of FL when the local data is non-i.i.d., highlighting the necessity of personalization. 
		To solve the non-i.i.d. challenges through personalization techniques, Meta-Learning methods are applied in FL \cite{Fallah2020, Jiang2019, Mikhail2019}.
		Smith \emph{et.al.} \cite{Smith2017} applied the multi-task learning (MTL) to FL and proposed the novel optimization method called MOCHA to solve the formulated MTL problem. The authors in \cite{Caldarola2021} addressed the statistical heterogeneity by clustering the agents and using the graph convolution networks to share knowledge across different clusters. The works in \cite{Collins2021} and \cite{Niko2022} apply the structural neural network architecture which consists of common layers across agents and the agent-specific layer for personalization. In addition to the non-i.i.d. challenges, some recent progress has been made on the communication-efficient implementation of FL in wireless communication system \cite{CEFL1, CEFL2, CEFL3, CEFL4}, through resource allocation \cite{CEFL1, CEFL2}, reducing communication cost per iteration \cite{CEFL3} and accelerating convergence \cite{CEFL4}.
		
		Different from FL, decentralized learning does not require a central node to collect and process all agents' information. Each agent shares the information with its neighboring nodes and aggregates the received messages locally, utilizing all agents' computational resources and alleviating the communication burden on the central server. Many recent treatises have paid attention to decentralized learning \cite{Lian2017, Scardapane2016, Jiang2017, Li2019, Balu2021, Liu2022}, which however are all based on the i.i.d. assumption. The authors in \cite{Esfandi2021, Tang2018, Lian2018, Ran2021, Jia2019, Pu2021} considered the condition of non-i.i.d. data distribution. \cite{Esfandi2021} proposed the Cross-Gradient Aggregation algorithm (CGA) to solve the statistical heterogeneity problem, which however takes high communication cost and they still intend to achieve a consensus model among all agents.
		{The authors in \cite{Ran2021, Jia2019, Pu2021} proposed algorithms based on gradient tracking, where the basic idea is to replace the local gradient with a tracker of global gradient.}
		On the other hand, the personalized model has been scarcely exploited in decentralized network. The personalization techniques for decentralized learning may have radical differences from those in FL, which need to take the network topology and local processing ability into consideration.
		The authors in \cite{valentina2020} leveraged a collaboration graph to describe the relationships among the users' tasks, which is learned alternately with the models. The proposed algorithm can obtain the personalized model for each agent. However, the alternate optimization procedure involving the graph learning may lead to high computation cost.
		Moreover, it requires the agents to communicate beyond their current direct neighbors in the communication network, which is impractical and may lead to high communication cost.
		
		{In this work, the non-i.i.d. challenge, the personalization needs and the robustness of decentralized network motivate us to develop personalized decentralized learning algorithm.}
		The learned model in each agent is expected to perform well w.r.t. the local data distribution, as is widely considered in practical scenarios, since the agent usually needs a personalized model to handle its local accessed data, rather than a poor-performed common global model.
		Inspired by the structural neural network proposed in \cite{Collins2021}, consisting of a shared data representation component and a unique head, we apply the partially-shared local model in each agent.
		We observe that the algorithm in \cite{Collins2021} trains the unique heads only with local data, without requiring any information from other agents. However, the non-i.i.d. data distributed in different agents usually has certain correlation, which can be exploited and utilized based on the topological structure formed by the agents.
		
		To achieve this goal, we treat each agent as one node and the node-specific parameters as its features. 
		We further deal with the topological structures of the parameters by introducing the graph neural network (GNN) \cite{Wu2019, Zhou2018, Meng2021}.
		{Numerous advanced models and architectures have been proposed 
			such as federated GNN in \cite{Meng2021} and minibatch graph convolutional networks \cite{Hong1}.
			It has also been widely applied in practical problems such as remote sensing and image processing problems \cite{Hong2, Hong3, Hong4}.}
		Specifically, the recently proposed graph attention network (GAT) \cite{GAT, Ryu2020} shows its effectiveness in specifying different importance for neighboring nodes, which can be utilized in the aggregation process.
		Besides, we observe that GAT is a parallelizable attention process without relying on any prior knowledge about the whole network, which can be used for decentralized implementations.
		Inspired by that, we propose to leverage the graph attention mechanism for decentralized learning, so as to pick up the effective information from other agents. Moreover, the different aggregation weights learned and assigned to various neighboring nodes can be utilized to reduce the communication cost, based on which we develop a communication-efficient training algorithm.
		
		Our contributions can be summarized as follows:
		\begin{itemize}
			\item{We propose GATTA to train the personalized model over network in non-i.i.d. condition, which fuses the graph attention mechanism into the decentralized learning. By jointly learning to specify the weights of different neighboring nodes in the training process, GATTA enables each agent to concentrate on the most relevant information received from its neighboring nodes.}
			\item{Based on the weight-learning mechanism of GATTA, we further design a communication-efficient GATTA (CE-GATTA) by skipping the transmission of less important information, which is characterized by the learned weights.}
			\item{We theoretically analyze the convergence properties of the proposed GATTA under given conditions, which provides a useful analytical approach for personalized learning. We show its convergence rate is $\mathcal{O}(\frac{1}{\sqrt{K}})$. Moreover, the range of the fusion parameter is derived, providing potential guidance on the parameter selection.}
			\item{Numerical experiments validate the superiority of GATTA and CE-GATTA compared with other methods in different datasets, including label distribution skew as well as feature distribution skew settings. Moreover, the results under different communication network topologies are evaluated and compared to show the effectiveness of GATTA more comprehensively. {Different local neural network architectures are simulated to show its broad applicability.} The communication cost is also investigated to show the communication efficiency of CE-GATTA.}
		\end{itemize}
		
		The rest of this paper is organized as follows. Section \ref{pre} describes the system model and the traditional D-SGD algorithm.
		In Section \ref{algo}, the partially-shared model among agents is introduced, based on which we develop the GATTA and CE-GATTA for personalized distributed learning. 
		Section \ref{theory} illustrates the assumptions and the convergence results of the proposed algorithm, where the range of the fusion parameter is derived. The simulation results are represented in Section \ref{experiment}, followed by the conclusion in Section \ref{conclusion}.
		
		{Note that this article significantly extends our previous work \cite{Tian2023} in several ways. 
			Firstly, we give the theoretical analysis of the convergence property, derive the convergence rate of GATTA and provide the range of the fusion parameter. 
			Secondly, we extend GATTA to a communication-efficient variant.
			Last but not the least, more experiments are conducted on different local neural network architectures and on the proposed CE-GATTA.
			We also compare the proposed methods with more state-of-the-art approaches on non-i.i.d. data.}
		
		\section{Preliminary}
		\label{pre}
		\subsection{System Model}
		Consider a multi-agent decentralized communication network, which can be represented by an undirected graph $\mathcal{G}=(\mathcal{V},\mathcal{E})$. 
		In $\mathcal{G}$, $\mathcal{V}=\{1,...,N\}$ denotes the set of $N$ distributed agents and $\mathcal{E}=\{\varepsilon_{ij}\}_{i,j \in \mathcal{V}} $ represents the set of communication links between any two adjacent agents. Let $\mathcal{N}_i$ denote the set of all neighboring agents connected with agent $i$ and we denote the number of agents in $\mathcal{N}_i$ by $d_i=|\mathcal{N}_i|$. The adjacency matrix of $\mathcal{G}$ is denoted by $\textbf{A}$, where $\textbf{A}(i,j)=1$ if $\varepsilon_{ij}\in\mathcal{E}$ and $\textbf{A}(i,j)=0$ otherwise. 
		
		Each agent $i\in\mathcal{V}$ has access to a local training dataset $\mathcal{D}_i=\{\bm{x}_s, \bm{y}_s\}_{s=1}^{n_i}$ {with the personal data distribution over some common feature space $\mathcal{X}$ and label space $\mathcal{Y}$.} 
		$n_i$ denotes the number of training samples in agent $i$. In the considered model, the data distributions in different agents are heterogeneous, known as non-i.i.d. data. 
		{
			In addition to the cooperative location problem introduced in Section \ref{sec1}, another example is the distributed natural language processing (NLP) problem, where each agent has a set of local users, whose distribution over words or expressions varies from one to another.
			The non-i.i.d. problem also arises in a distributed sensing system, where the agents collaboratively sense some signal. Usually, the observed signal of each agent has personalized degradation, noise effects, or variabilities \cite{Hong5}.
		}
		
		Let $f_i$ denote the loss function corresponding to agent $i$ and the global loss function of the whole network is:
		\begin{equation}
			\label{opt_sys}
			\min F(\textbf{V}) := \frac{1}{N}\sum\nolimits_{i=1}^{N}f_i(\bm{v}_i),
		\end{equation}
		where $\bm{v}_i$ denotes the model parameters of node $i$.
		Particularly, in a supervised learning setting, $f_i(\bm{v}_i)$ stands for the expected loss over the local data distribution of agent $i$ and is defined as $f_i(\bm{v}_i) := \mathbb{E}_{\mathcal{D}_i}[l_i(\bm{v}_i; \bm{x}_s, \bm{y}_s)]$, where $l_i(\bm{v}_i; \bm{x}_s, \bm{y}_s)$ measures the error in predicting the label $\bm{y}_s$ given the input $\bm{x}_s$ and the model parameters $\bm{v}_i$.
		
		{Conventional consensus-based methods, such as FL or decentralized stochastic gradient descent, aim at minimizing the global loss function in (\ref{opt_sys}) with consensus constraints $\bm{v}_1=\bm{v}_2=\cdots=\bm{v}_N$. However this approach performs poorly in the heterogeneous settings and personalized learning tasks due to different distributions of $\mathcal{D}_i$.
			To this end, the optimization problem in (\ref{opt_sys}) is proposed without consensus constraints. 
			Observing (\ref{opt_sys}), it seems that each agent can learn its own model independently, without communicating with others.
			However, in many typical distributed learning settings, the number of local samples is small and cannot give accurate estimation of the expectation in $f_i(\bm{v}_i)$.
			Thus, it cannot promise solutions with small expected risk by training completely locally. 
			In this sense, the collaboration among agents is necessary and through exploiting the available information from other agents, the local models can be well improved.
		}
		
		Before developing our algorithm, we introduce the decentralized stochastic gradient descent training method applied in distributed learning to achieve consensus among the learning model of different agents. 
		
		\subsection{Decentralized SGD Method}
		A widely-used method for distributed training is the decentralized stochastic gradient descent (D-SGD) {\cite{Lian2017}}, which averages the model parameters from neighboring agents in each iteration. D-SGD is a {simple} yet efficient algorithm when applied to learn a common model for agents. 
		Specifically, in the $k$-th round, each agent updates the model parameters $\bm{v}_i$ in two steps: first carrying out one epoch of local stochastic gradient descent {(SGD)} \cite{Bottou2018} to obtain an intermediate variable and then aggregating the obtained neighboring agents' parameters to complete the update. Here one epoch refers to a few steps of stochastic gradient descent (SGD), which walks through all the local training samples.
		As discussed in the introduction, the simple mechanism of averaging fails to exploit the correlated information among them, which may instead lead to worse performance in the non-i.i.d. case. 
		To this end, we propose an algorithm in the following section, which can intelligently aggregate information from neighboring nodes.
		
		\section{Distributed Learning with Graph-Attention-Based Personalization}
		\label{algo}
		In this section, we develop the graph attention-based personalized training algorithm for distributed learning over a network.
		We first present the partially-shared local model in each client, and then illustrate the graph attention-based aggregation procedure, which can learn to exploit the useful information from other agents. The attention-based distributed training algorithm GATTA for personalized learning is summarized after that. 
		Thirdly, based on the proposed GATTA, we develop the communication-efficient GATTA.

		\subsection{The Partially-Shared Local Model}
		\renewcommand{\thefootnote}{(a)}
		
		\begin{figure*}[!htp]
			\centering
			\vspace{-0.5cm}
			\includegraphics[width=0.95\textwidth]{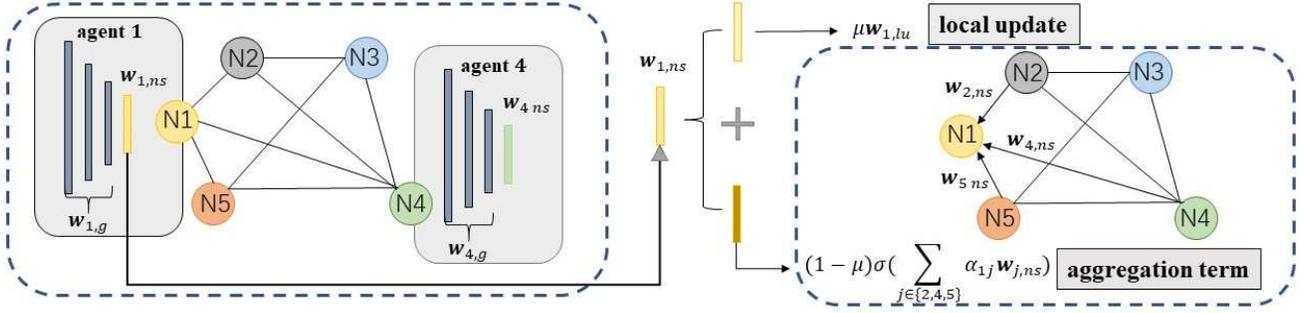}
			\caption{The system model of GATTA: The left side represents the decentralized learning framework, where the partially-shared model is shown in the gray box. The subscript $g$ indicates the global model part and the subscript $ns$ stands for node-specific layer. The right side represents the two components of the node-specific parameters.}
			\vspace{-0.5cm}
			\label{model}
		\end{figure*}
		
		We are motivated by the work in \cite{Collins2021}, which separates the local deep neural network into the common global layers and the personalized head unique for each agent. This insight comes from the traditional machine learning which suggests that the heterogeneous data may share a global representation despite having different labels. 
		Inspired by that, we apply the partially-shared neural network architecture as the local model in each agent. Specifically, in the local neural network, the front layers, mapping the input into lower dimensions, are shared among all agents, while the last one layer performs as the node-specific part and is unique for each agent \footnote{In this work, we only consider the last one layer as the node-specific part for DNN. The proposed scheme can be easily generalized to multi-layer conditions with proper design for other network architectures.}. The partially-shared model is shown in Fig. \ref{model}.
		
		With the partially-shared model, after one epoch of local training, each agent transmits the model parameters to its neighboring agents, including the global ones $\bm{w}_{i,g}$ as well as the node-specific ones $\bm{w}_{i, ns}$. To achieve the consensus of the global model, the efficient D-SGD method {\cite{Lian2017}} is applied to update $\bm{w}_{i,g}$ in each round. 
		\cite{Collins2021} proposed to train the node-specific parameters only using local data set, without utilizing information from other agents. However, as mentioned before, this is inefficient especially in the condition of small number of training samples.
		To this end, we design an aggregation procedure for the node-specific parameters based on the graph attention mechanism.
		
		\subsection{The Graph Attention-Based Aggregation Method}\label{s_GAT}
		{The Graph Attention Network was first proposed in \cite{GAT}, which is a novel neural network architecture extending conventional neural networks to deal with graph-structured data.
			It can be applied to various tasks in graph domain, such as node classification and regression.
			Through utilizing self-attention layers, GAT can automatically configure different weights for different neighboring nodes without requiring any pre-defined weight matrix.
		}
		
		Inspired by this automatic weight configuration mechanism and the fact that a decentralized communication network can be viewed as a graph, we propose a graph attention based network for personalized model aggregation, aiming to dynamically exploit the effective information from other nodes to boost the local model. 
		Each agent can be treated as one node in the graph, while its node-specific parameters are treated as local features and can be aggregated according to the graph attention mechanism. It can be implemented in a node-wise manner, which is {suitable} for the decentralized architecture.
		{Following we will talk about how to utilize the graph attention mechanism in distributed personalized learning over networks.}
		
		
		Consider a node $i$, representing the agent $i$, with its node-specific parameters $\bm{w}_{i, ns}\in\mathbb{R}^{F}$.
		In the $k$-th round, after receiving the neighboring nodes' parameters in the last round, the input of the attention mechanism for each node is a set of node features $\{\bm{w}_{j,ns}^{(k-1)}\}_{j\in\mathcal{N}_i{\cup \{i\}}}$, consisting of its local node-specific parameters as well as those from neighboring nodes. 
		{Here the symbol $\mathcal{N}_i{\cup \{i\}}$ denotes the union of $\mathcal{N}_i$ and the local node.}
		A locally shared attention mechanism $a_i(\cdot): \mathbb{R}^{F}\times \mathbb{R}^{F} \to \mathbb{R}$ is applied in the concatenation of $\bm{w}_{i, ns}^{(k-1)}$ and $\bm{w}_{j, ns}^{(k-1)}, j\in\mathcal{N}_i$ to compute the attention coefficient:
		\begin{equation}
			\label{eijk}	
			e_{ij}^{(k)}=a_i(\bm{w}_{i, ns}^{(k-1)}||\bm{w}_{j, ns}^{(k-1)}), j\in\mathcal{N}_i,
		\end{equation}
		where $||$ denotes the concatenation operation.
		The attention coefficient $e_{ij}^{(k)}$ indicates the importance of node $j$'s parameters to node $i$.
		A softmax operation is {conducted on $e_{ij}^{(k)}$} for coefficient normalization across all neighboring nodes. 
		{Thus we have:
			\begin{equation*}
				\alpha_{ij}^{(k)}=\text{softmax}_j(e_{ij}^{(k)})=\frac{\exp(e_{ij}^{(k)})}{\sum_{l\in\mathcal{N}_i}\exp(e_{il}^{(k)})}.
		\end{equation*}}
		In practical implementations, {to compute the attention coefficient $e_{ij}^{(k)}$ in (\ref{eijk})}, the attention mechanism $a_i(\cdot)$ is a single 1-dimensional convolution layer, parameterized by the weight parameters $\bm{\beta}_i\in\mathbb{R}^{2F}$. Applying the activation function $\sigma_G[\cdot]$, the aggregation weights computed by the attention mechanism can be expressed as:
		\begin{equation}
			\label{att_wei}
			\alpha_{ij}^{(k)}=\frac{\exp\Big(\sigma_G\big[{\bm{\beta}_i^{(k)}}^T(\bm{w}_{i, ns}^{(k-1)}||\bm{w}_{j, ns}^{(k-1)})\big]\Big)}{\sum_{l\in\mathcal{N}_i}\exp\Big(\sigma_G\big[{\bm{\beta}_i^{(k)}}^T(\bm{w}_{i, ns}^{(k-1)}||\bm{w}_{l, ns}^{(k-1)})\big]\Big)}.
		\end{equation}
		Note that the original activation function $\sigma_G[\cdot]$ applied in \cite{GAT} is LeakyReLU, while ELU is used as the activation function in our model for the sake of smoothness.
		
		
		Meanwhile, considering that the aggregation of the neighboring nodes' parameters may be insufficient, we utilize another intermediate local update parameter term, denoted by $\bm{w}_{i, lu}$.
		The aggregation model for the node-specific parameters can thus be formulated as:
		\begin{equation}
			\label{fuse2}
			\bm{w}_{i, ns}^{(k)} \gets \underbrace{\mu \bm{w}_{i, lu}^{(k)}}_{\text{local update}} + \underbrace{(1-\mu) \sigma\Big( \sum_{j\in\mathcal{N}_i} \alpha_{ij}^{(k)}\bm{w}_{j, ns}^{(k-1)}\Big)}_{\text{aggregation term}},
		\end{equation}
		where $\sigma$ is an activation function.
		Here $\mu$ is the fusion parameter to balance the local update and the aggregation of parameters from neighboring nodes. We derive the range of $\mu$ in Theorem \ref{t3} in the next section, whose lower bound increases as the degree of non-i.i.d. becomes large.
		Note that in (\ref{fuse2}), the variables updated in the back propagation include the weight parameters $\bm{\beta}_i$ in the attention mechanism as well as the intermediate local update parameters $\bm{w}_{i, lu}$. 
		
		With the attention-based mechanism, the graph-attention-based personalized training algorithm for decentralized learning, termed as GATTA, is summarized as Algorithm \ref{alg1}. Here $\tilde{\textbf{A}}$ denotes the weight matrix for aggregating the global model parameters. 
		
		\begin{algorithm}
			{\small
				\caption{\textbf{Graph-Attention Based Training Algorithm (GATTA)}}
				\label{alg1}
				\For{each node $i\in\mathcal{V}$ [in parallel]}
				{
					\textbf{Initialize} all the parameters in the network and set $k=0$.\\
					\textbf{Initialize} the parameters of neighboring nodes. 
					\\
				}
				\While{not converged}
				{
					$k\gets k+1$.\\
					\For{each node $i\in\mathcal{V}$ [in parallel]}
					{
						\For{all local training samples $\{\bm{x}_s, \bm{y}_s\}\in\mathcal{D}_i$ randomly}
						{
							\textbf{Input} the training samples and \textbf{compute} the loss with the current model parameters.\\
							\textbf{Back propagate} the gradients and \textbf{Update} the model parameters.\\ 
						}
						{ \textbf{Obtain} $\bm{w}_{i, g}^{(k-\frac{1}{2})}$, $\bm{w}_{i, lu}^{(k)}$, $\bm{\beta}_i^{(k)}$ after the local SGD.} \\
						\textbf{Calculate} the node-specific parameters $\bm{w}_{i, ns}^{(k)}$ according to (4).\\
						\textbf{Transmit} the model parameters $\bm{w}_{i, g}^{(k-\frac{1}{2})}$ and $\bm{w}_{i, ns}^{(k)}$ to the neighboring nodes.\\
					}
					\For{each node $i\in\mathcal{V}$ [in parallel]}
					{
						\textbf{Update} the global model parameters $\bm{w}_{i, g}^{(k)}\gets\sum_{j\in\mathcal{N}_i\cup \{i\}}\tilde{\textbf{A}}(i,j)\cdot\bm{w}_{j, g}^{(k-\frac{1}{2})}$.\\
					}
				}
			}
		\end{algorithm}
		
		{For a given neural network architecture in each agent, such as AlexNet, we denote the number of its parameters by $N_{v} = N_{wg}+N_{wlu}$, where $N_{wg}$ and $N_{wlu}$ represent the number of parameters in the global model part and node-specific part.
			Then in one iteration, the number of parameters to be updated in D-SGD is $N_{v}$ for one agent. The number in D-SGD with gradient tracking (GT-DSGD) \cite{Ran2021} is $2N_{v}$.
			For GATTA, the parameters to be updated include $\bm{w}_{i, g}$, $\bm{w}_{i, lu}$ and $\bm{\beta}_i$, thus the total number is $N_{v}+2N_{wlu}$.
			Since the node-specific layer takes a small part in the neural network, we have $N_v\le N_{v}+2N_{wlu}\le2N_v$.}
		
		\subsection{Communication-Efficient GATTA for Distributed Learning}
		\label{CE_GATTA}
\begin{figure}[!htp]
			\centering
			\includegraphics[width=0.4\textwidth]{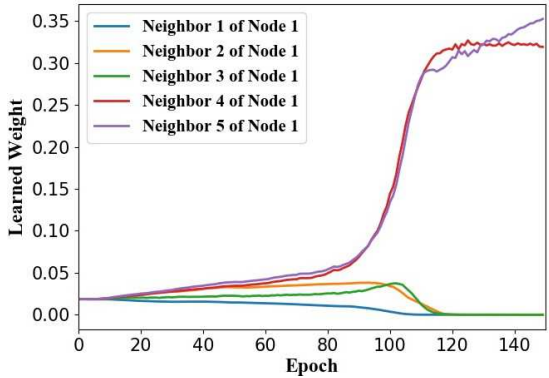}
			\caption{Different neighboring nodes' weights}
			\label{weight}
		\end{figure}

		In the training process of GATTA, each agent adaptively decides how to fuse the node-specific parameters from its neighboring nodes through learning the aggregation weights. For each node, different neighboring nodes with various data distribution may have different impact on it, leading to diverse weights, especially in label distribution condition \cite{Li2021}. Then it naturally comes to us that in the training process, each node can stop receiving node-specific parameters from those neighboring nodes which have little positive impact on it with small weights.
		To better illustrate this, we plot Fig. \ref{weight} following the same setting as the first experiment in Section \ref{experiment}. We take an arbitrary node for representation and show the learned weights of its selected $5$ different neighboring nodes. 
		
		It can be observed that with the iteration going on, the weights of some neighboring nodes reduce to small values, which means those nodes have little positive impact on the local one. 
		Motivated by this observation, we further design a communication-efficient GATTA (CE-GATTA). 
		Specifically, we set the weight threshold $\tau_i$. When the learned weight of $j$-th neighboring nodes is less than $\tau_i$, i.e., $\alpha_{ij}<\tau_i$, the $j$-th neighboring node stops to transfer its node-specific parameters to node $i$. 
		The aggregation model can thus be reformulated as:
		\begin{equation}
			\label{fuse3}
			\bm{w}_{i, ns}^{(k)} \gets{\mu \bm{w}_{i, lu}^{(k)}} + (1-\mu) \sigma\Big( \sum_{j\in\mathcal{N}_{c,i}^{(k)}} \alpha_{ij}^{(k)}\bm{w}_{j, ns}^{(k-1)}\Big),
		\end{equation}
		where $\mathcal{N}_{c,i}^{(k)}$ denotes the set of the selected neighboring nodes which need to transfer their node-specific parameters in the $k$-th iteration.
		Note that those neighboring nodes outside $\mathcal{N}_{c,i}$ only stop transmitting their node-specific parameters, rather than stopping transmitting the global model's parameters. 
		This may lead to higher communication cost per epoch compared with totally stopping the communication from this node. However, it is necessary to share the global model part so as to guarantee convergence rate with the information from other nodes. 
		
		Since the aggregation weight of the removed information is small, such reduction of communication may have little impact on the convergence performance of the algorithm compared with original GATTA.
		{Moreover, the whole information still flows over the connected communication network and CE-GATTA can adjust the aggregation weights to better fuse the information, as also indicated by experimental results.}
		Consequently, under the proper choice of $\tau_i$, CE-GATTA may have similar performance w.r.t. convergence rate and resultant accuracy compared with the original GATTA.
		Experimental results in Section \ref{experiment} show a much faster convergence rate of GATTA compared with D-SGD.
		Thus, through reducing the communication per iteration as well as reducing the total communication rounds, the communication cost of the system can be highly saved. The communication-efficient GATTA can be summarized as Algorithm \ref{alg2}. 
		
		{Note that the amount of communication cost saved by CE-GATTA is relevant to the value of the threshold $\tau_i$.
			When $\tau_i$ is relatively small, the number of removed communication nodes is small and has little impact on the convergence curve.
			However, when $\tau_i$ becomes large enough, the convergence curve of CE-GATTA may become lossy and it takes more iterations to aggregate the information so as to achieve the same accuracy as GATTA. 
			Consequently, the total communication cost maybe not necessarily saved. 
			In this sense, there exists a trade-off between the number of iterations and the communication cost saved in each iteration, which is determined by $\tau_i$ and eventually affects the overall communication cost.
			On the other hand, the performance of CE-GATTA maybe also influenced by the non-i.i.d. properties of the data distribution.
			For example, for the feature distribution skew condition in Section \ref{experiment}, the learnt aggregation weight has small divergence among neighboring nodes, where the effect of CE-GATTA is weakened.
            In contrast, the divergence of the learnt weights in label distribution skew is large as shown in Fig. \ref{weight}, and CE-GATTA has remarkable performance on saving the communication cost.
		}
		
		\begin{algorithm}
			{\small
				\caption{\textbf{Communication-Efficient GATTA (CE-GATTA)}}
				\label{alg2}
				\For{each node $i\in\mathcal{V}$ [in parallel]}
				{
					\textbf{Initialize} all the parameters in the network and set $k=0$, $\mathcal{N}_{c,i}=\mathcal{N}_i$.\\
					\textbf{Initialize} the parameters of neighboring nodes. 
					\\
				}
				\While{not converged}
				{
					$k\gets k+1$.\\
					\For{each node $i\in\mathcal{V}$ [in parallel]}
					{
						\For{all local training samples $\{\bm{x}_s, \bm{y}_s\}\in\mathcal{D}_i$ randomly \footnotemark{}}
						{
							\textbf{Input} the training samples and \textbf{compute} the loss with the current model parameters.\\
							\textbf{Back propagate} the gradients and \textbf{Update} the model parameters.\\ 
						}
						{\textbf{Obtain} $\bm{w}_{i, g}^{(k-\frac{1}{2})}$, $\bm{w}_{i, lu}^{(k)}$, $\bm{\beta}_i^{(k)}$ after the local SGD and \textbf{Calculate} $\{\alpha_{ij}^{(k)}\},j\in\mathcal{N}_{c,i}$ according to (3).}\\
						\textbf{Calculate} the node-specific parameters $\bm{w}_{i, ns}^{(k)}$ according to (5).\\
						{\textbf{For all }$\alpha_{ij}^{(k)}<\tau_i$, remove $j$ from $\mathcal{N}_{c,i}^{(k-1)}$ and get $\mathcal{N}_{c,i}^{(k)}$.\\
							\textbf{If} $\mathcal{N}_{c,i}^{(k)}=\emptyset $, then $\mathcal{N}_{c,i}^{(k)} \leftarrow \mathcal{N}_{c,i}^{(k-1)}$ .\\
							\textbf{Inform} those neighboring nodes outside  $\mathcal{N}_{c,i}^{(k)}$ to stop transmitting $\bm{w}_{j, ns}$.\\}
						
						\textbf{Transmit} the model parameters $\bm{w}_{i, g}^{(k-\frac{1}{2})}$ to all neighboring nodes.\\
						\textbf{Transmit} $\bm{w}_{i, ns}^{(k)}$ to the needed neighboring nodes.\\
					}
					\For{each node $i\in\mathcal{V}$ [in parallel]}
					{
						\textbf{Update} the global model parameters $\bm{w}_{i, g}^{(k)}\gets\sum_{j\in\mathcal{N}_i\cup \{i\}}\tilde{\textbf{A}}(i,j)\bm{w}_{j, g}^{(k-\frac{1}{2})}$.\\
					}
				}
			}
		\end{algorithm}

		\section{Theoretical Results}\label{theory}
		We denote the updatable parameters of the local neural network in node $i$ by $\bm{v}_i$, which is the concatenation of the global parameters $\bm{w}_{i, g}$, the local update parameters $\bm{w}_{i,lu}$, and the attention parameters $\bm{\beta}_i$.
		In the $k$-th round, the update rules of the parameters can be written as follows:
		\begin{equation}
			\label{e1}
			\bm{w}_{i, g}^{(k-\frac{1}{2})}=\bm{w}_{i, g}^{(k-1)}-\eta\sum_{t=0}^{T-1}g_{i,g,t}^{(k-1)}=\bm{w}_{i, g}^{(k-1)}-\eta\Delta_{i,g}^{(k-1)},
		\end{equation}
		\begin{equation}
			\label{e4}
			\bm{w}_{i, g}^{(k)}=\sum\nolimits_{j\in\mathcal{N}_i\cup \{i\} }\tilde{\textbf{A}}(i,j)\bm{w}_{j, g}^{(k-\frac{1}{2})},
		\end{equation}
		\begin{equation}
			\label{e2}
			\bm{w}_{i, lu}^{(k)}=\bm{w}_{i, lu}^{(k-1)}-\eta\sum_{t=0}^{T-1}g_{i,lu,t}^{(k-1)}=\bm{w}_{i, lu}^{(k-1)}-\eta\Delta_{i,lu}^{(k-1)},
		\end{equation}
		\begin{equation}
			\label{e3}
			\bm{\beta}_{i}^{(k)}=\bm{\beta}_{i}^{(k-1)}-\eta\sum_{t=0}^{T-1}g_{i,b,t}^{(k-1)}=\bm{\beta}_{i}^{(k-1)}-\eta\Delta_{i,b}^{(k-1)},
		\end{equation}
		where $\eta$ denotes the learning rate at the $k$-th communication round and $T$ is the number of stochastic gradient descent (SGD) steps in one epoch. After the $k$-th communication round, $g_{i,g,t}^{(k)}$ denotes the gradient w.r.t. the global parameters $\bm{w}_{i, g}$ in the $t$-th SGD step. Likewise, the subscript $lu$ in (\ref{e2}) and $b$ in (\ref{e3})  respectively represents $\bm{w}_{i,lu}$ and $\bm{\beta}_i$. $\Delta$ denotes the accumulated gradients after one epoch of SGD. 
		Define the gradient of the local objective w.r.t. any parameter $\bm{w}$ as $\nabla f_i(\bm{w})$, then we have $g_{i,g,t}^{(k)} = \nabla f_i(\bm{w}_{i,g,t}^{(k)}, \xi_{i,t})$, where $\xi_{i,t}$ denotes the data samples in the $t$-th SGD step.
		In the $k$-th iteration, we define the averaged global parameters among agents as $\bar{\bm{w}}_{g}^{(k)}$. 
		Meanwhile, we use the matrix $\textbf{W}_{g}$ to represent the  matrix form of the global parameters for all agents, i.e., $\textbf{W}_{g}=[\bm{w}_{1,g}, \bm{w}_{2,g}, ..., \bm{w}_{N,g}]$, where $\bm{w}_{i,g}$ is a column vector here. Likewise, the matrix form of $\Delta_{i,g}$ for all agents can be denoted by $\Xi_{g}\triangleq[\Delta_{1,g},..., \Delta_{N,g}]$.
		To this end, based on the update rule in (\ref{e1}) and (\ref{e4}), we can obtain the following update rule in matrix form.
		\begin{equation}
			\label{e5}
			\textbf{W}_{g}^{(k)}=(\textbf{W}_{g}^{(k-1)}-\eta\Xi_{g}^{(k-1)})\tilde{\textbf{A}}.
		\end{equation}
	
		Before presenting our theoretical findings, we make the following assumptions, where the expectations are taken over the randomness in stochastic gradients.
		\newtheorem{assumption}{Assumption}
		\begin{assumption} (Spectral Gap)
			\label{a6}
			The aggregation weight matrix $\tilde{\textbf{A}}$ for the global model parameters is a symmetric doubly stochastic matrix.  Denote its eigenvalues by $1=|\lambda_1|>|\lambda_2|\ge\cdots\ge|\lambda_N|\ge0$. We further assume the spectral gap $1-\rho\in(0, 1]$, where $\rho=|\lambda_2|\in(0, 1]$.
		\end{assumption}
		
		\begin{assumption} (Smoothness)
			\label{a1}
			The local objective functions $f_i$ are $L$-smooth for the parameters $\bm{v}_i$ of each node $i\in\mathcal{V}$, i.e.,  
			\begin{equation}
				f_i(\bm{v}_i)\le f_i(\bm{v}_i')+\nabla f_i(\bm{v}_i')^T(\bm{v}_i-\bm{v}_i')+\frac{L}{2}\|\bm{v}_i-\bm{v}_i'\|^2_2.
			\end{equation}
		\end{assumption}
		
		\begin{assumption}(Unbiased Local Gradient Estimator) 
			\label{a2}
			For each node $i\in\mathcal{V}$ and $\bm{w}_i\in\{\bm{w}_{i, g}, \bm{w}_{i, lu}, \bm{\beta}_{i}\}$, the local gradient estimator is unbiased, i.e., $\mathbb{E}[g_{i,t}]=\nabla f_i(\bm{w}_{i,t})$, in the $t$-th gradient descent step.
		\end{assumption}
		
		\begin{assumption}(Bounded Local Variance)
			\label{a3}
			There exist scalar $\chi>0$ such that for each node $i\in\mathcal{V}$ and $\bm{w}_i\in\{\bm{w}_{i, g}, \bm{w}_{i, lu}, \bm{\beta}_{i}\}$, the variance of local gradient estimator is bounded by $\mathbb{E}[\|g_{i,t}-\nabla f_i(\bm{w}_{i,t})\|]\le \chi$.
		\end{assumption}
		
		\begin{assumption} (Degree of Non-i.i.d.)
			\label{a5}
			There exists scalar $\kappa\ge 0$ for each node $i\in\mathcal{V}$ such that for the global parameters,
			\begin{equation}
				\frac{1}{N}\sum_{i\in\mathcal{V}}\mathbb{E}\Big\|\nabla f_i(\bm{w}_{g})-\frac{1}{N}\sum_{j\in\mathcal{V}}\nabla f_j(\bm{w}_{g})\Big\| \le \kappa.
			\end{equation}
		\end{assumption}
		
		\begin{assumption} (Bounded Gradients)
			\label{a4}
			There exists scalar $G>0$ such that for each node $i\in\mathcal{V}$ and any $\bm{w}_i\in\{\bm{w}_{i, lu}, \bm{\beta}_{i}, \bm{w}_{i, ns}\}$,
			\begin{equation}
				\label{a4_1}
				\|\nabla f_i(\bm{w}_i)\|_2^2\le G.
			\end{equation}
			Further, the gradients of the activation functions satisfy
			\begin{equation}
				\label{a4_2}
				\|\sigma'\|_2^2\le 1.
			\end{equation}
		\end{assumption}
		
		Among all assumptions, Assumption \ref{a1} for the local objective function is standard, which also restricts the activation functions to be smooth. The commonly used activation functions such as ELU, sigmoid and Tanh all satisfy this assumption. Assumption \ref{a5} limits the non-i.i.d. degree through the gradients of the global parameters $\bm{w}_{g}$.
        Assumptions \ref{a6}-\ref{a5} are commonly used and can be widely found in \cite{Bottou2018, Fallah2020, Li2019, Lian2017, Dinh2020, Haibo2021, Tian2021}. 
        We further make Assumption \ref{a4} to simplify the analysis in Theorem \ref{t3}, where equation (\ref{a4_2}) can be easily satisfied by most common activation functions including ELU, sigmoid and Tanh.
		
		Here one key difference of our analysis from others in standard D-SGD is that we take the node-specific parameters into consideration.
		Specifically, in our analysis, the performance is evaluated under the \textit{averaged} global model parameters among agents $\bar{\bm{w}}_g$, together with the \textit{personalized} individual parameters including local update parameters $\bm{w}_{i,lu}$ and attention parameters $\bm{\beta}_{i}$. 
		This is rational because the parameters in the global model part are updated following the standard D-SGD to achieve consensus among agents, while the other parameters are updated locally with personalization. The performance of D-SGD has been analyzed in \cite{Lian2017,  Li2019} based on averaged parameters. Different from them,  we additionally consider the node-specific model part and combine these two kinds of parameters.
		Given the above assumptions, we have the following lemmas, where the expectation is over the local data samples.
		
		\newtheorem{lemma}{Lemma}
		\begin{lemma}
			\label{l1}
        	Denote the variable value of $\bm{w}_{i}$ in the $t$-th SGD step by $\bm{w}_{i, t}$. Under Assumption \ref{a3}, we have
			\begin{equation}
				\label{l1_1}
				\mathbb{E}[\|\Delta_{i}^{(k)}\|_2^2]\le T\cdot\chi^2+\mathbb{E}[\|\sum_{t=0}^{T-1}\nabla f_i(\bm{w}_{i, t}^{(k)})\|_2^2],
			\end{equation}
			for all $\bm{w}_i\in\{\bm{w}_{i, g}, \bm{w}_{i, lu}, \bm{\beta}_{i}\}$, and $\Delta_i\in\{\Delta_{i, g}, \Delta_{i, lu}, \Delta_{i,b}\}$ respectively.
		\end{lemma}
		
		\begin{lemma}
			\label{l2}
			For any learning rate satisfying $\eta<\frac{1}{4TL}$, we have the following results:
			\begin{equation}
				\label{l5_1}
				\mathbb{E}[\|\bm{w}_{i,t}-\bm{w}_i\|_2^2]\le4T\eta^2\chi^2+16T^2\eta^2\|\nabla f_i(\bm{w}_i)\|_2^2.
			\end{equation}
		\end{lemma}

		\begin{lemma}
			\label{l3}
			For any $i\in\mathcal{V}$, $\bm{w}_i\in\{\bm{w}_{i, lu}, \bm{\beta}_{i}\}$ and $\Delta_i\in\{\Delta_{i, lu}, \Delta_{i,b}\}$ respectively, we have
			\begin{equation}
				\label{e6}
				\begin{aligned}
					& \mathbb{E}\Big[-\eta\nabla f_{i}(\bm{w}_{i}^{(k-1)})^T\Delta_i^{(k-1)}+\frac{1}{2}\eta^2L\|\Delta_i^{(k-1)}\|^2_2 \Big] \\
					& \le-cT\eta\|\nabla f_{i}(\bm{w}_{i}^{(k-1)})\|_2^2+\frac{\eta^2TL}{2}(1+4\eta TL)\chi^2,
				\end{aligned}
			\end{equation}
			where $c$ is a constant satisfying $0<c<\frac{1}{2}-8\eta^2T^2L^2$.
		\end{lemma}

		\begin{lemma}
			\label{l4}
			For the global parameters $\bm{w}_{g}$, we have
			\begin{equation}
				\label{el4}
				\mathbb{E}\Big\|\nabla F(\bar{\bm{w}}_g^{(k)})-\nabla F(\bm{w}_g^{(k)})\Big\|^2_2 \le \frac{L^2}{N}\sum_{i\in\mathcal{V}}\mathbb{E}\|\bar{\bm{w}}_g^{(k)}-\bm{w}_{i,g}^{(k)}\|_2^2, 
			\end{equation}
            where $\nabla F(\bar{\bm{w}}_g^{(k)})\triangleq\frac{1}{N}\sum_{i\in\mathcal{V}}\nabla f_i(\bar{\bm{w}}_{g}^{(k)})$ and $\nabla F(\bm{w}_g^{(k)})\triangleq\frac{1}{N}\sum_{i\in\mathcal{V}}\nabla f_i(\bm{w}_{i,g}^{(k)})$
		\end{lemma}

            \begin{lemma}
			\label{l5}
			For the averaged global parameters $\bar{\bm{w}}_{g}$, we have
			\begin{equation}
				\label{el5}
					\bar{\bm{w}}_g^{(k)}-\bar{\bm{w}}_g^{(k-1)} = -\frac{\eta}{N} \sum_{i\in\mathcal{V}}\Delta_{i, g}^{(k-1)} \triangleq-\eta\bar{\Delta}_g^{(k-1)}. \notag
			\end{equation}
		\end{lemma}

		The proofs of the Lemmas can be found in Appendix A.
		As we talked before, the performance is measured under averaged $\bar{\bm{w}}_g$ and individual $\bm{w}_{i,lu}$, $\bm{\beta}_{i}$. Thus we define the partially-shared parameters in agent $i$ as $\tilde{\bm{v}}_i$, which is the concatenation of  $\bar{\bm{w}}_g$, $\bm{w}_{i,lu}$ and $\bm{\beta}_{i}$. Additionally, we define the concatenation of individual parameters $\bm{w}_{i,lu}$, $\bm{\beta}_{i}$ as $\bm{v}_{i,ns}$ and $\nabla F(\bm{v}_{ns}) \triangleq\frac{1}{N}\sum_{i\in\mathcal{V}}\nabla f_i({\bm{v}}_{i,ns})$. 
	  The product of multiple weight matrices for global parameters is denoted by $\bar{\bar{\textbf{A}}}_{s, k-1} = \prod_{l=s}^{k-1}\tilde{\textbf{A}}$.  $\textbf{Q}=\frac{1}{N}\textbf{1}_N\textbf{1}_N^T$ and $\rho_{s, k-1}=\|\bar{\bar{\textbf{A}}}_{s, k-1}-\textbf{Q}\|$. Additionally, we also make the following definitions.
        \begin{align}
		&A_K=\frac{1}{K}\sum_{k=1}^{K}\sum_{s=1}^{k-1}\rho_{s, k-1}^2,\quad B_K=\frac{1}{K}\sum_{k=1}^{K}\Big(\sum_{s=1}^{k-1}\rho_{s, k-1}\Big)^2,\notag \\
		& C_K = \max_{s\in[K-1]}\sum_{k=s+1}^{K}\rho_{s, k-1}\Big(\sum_{l=1}^{k-1}\rho_{l, k-1}\Big). \notag
		\end{align}
    Then based on the lemmas and definitions above, we give the convergence property of the proposed GATTA method as Theorem \ref{t1} and Corollary \ref{c1}. 
		
		\newtheorem{theorem}{Theorem}
		\begin{theorem}
			\label{t1}
            Provided that $\eta<\min\{\frac{1}{24TL}, \frac{1}{32TL\sqrt{C_K}}\}$, under Assumptions 1-5 made above, the iterates of GATTA algorithm satisfy the following inequality:
			\begin{align}
				\label{e8}
            &\min_{k\in[K]}\mathbb{E}\big[\|\nabla F({\bm{v}}_{ns}^{(k)})\big\|_2^2+\|\nabla F(\bar{\bm{w}}_{g}^{(k)}) \|_2^2\big]\le \frac{F_0-F_*}{cTK\eta}+\Phi, \notag
			\end{align}
			where $F_0$ denotes initial value of the objective $F(\tilde{\textbf{V}})$ and $F_*$ denotes its optimal value.  
    \begin{align}
    \Phi= \frac{1}{c}\Big\{&\eta L(1+4\eta TL)\chi^2 + \notag\\
& \frac{1}{N}\big[\eta L(4\kappa^2T+\chi^2) + 6T\eta^2\chi^2L^2\big] + \notag\\
  & 64\eta^2TL^2(A_K\chi^2+B_KT( \kappa^2+T\eta^2\chi^2L^2))\Big\},\notag
   \end{align}
   $c$ is a constant satisfying $0<c<\frac{1}{2}-8\eta^2T^2L^2$, and $A_K, B_K, C_K$ are defined as above.
		\end{theorem}
		Its proof can be found in Appendix B.
		Based on Theorem \ref{t1}, we have the following convergence rate for GATTA as Corollary \ref{c1}.
		
		\newtheorem{corollary}{Corollary}
		\begin{corollary}
			\label{c1}
			Let the learning rate $\eta=\frac{m}{\sqrt{K}}$, where $m$ is a constant such that $\eta<\min\{\frac{1}{24TL}, \frac{1}{32TL\sqrt{C_K}}\}$, then the convergence rate for GATTA is $\mathcal{O}(\frac{1}{\sqrt{K}})$.
		\end{corollary}
		
		Finally, we provide the limited range of the fusion parameter $\mu$ in (\ref{fuse2}) as the following Theorem \ref{t3}. We denote $\mu$ by $\mu_i^{(k)}$ to associate with a specific node $i$ in the $k$-th round for better clarification.
		
		\begin{theorem}
			\label{t3}
			Denote the gradient value of $\sigma_G({x}_j)$ by $\sigma_{G,j}'$, where ${x}_j\triangleq {\bm{\beta}_i^{(k-1)}}^T(\bm{w}_{i,ns}^{(k-1)}||\bm{w}_{j,ns}^{(k-1)})$. Define
			\begin{equation*}
				\begin{aligned}
					& D_i^{(k)}\triangleq \\ 
					&\big[\sum_{j\in\mathcal{N}_i}\big\|\bm{w}_{j,ns}^{(k-1)}\|_2^2\big]\cdot\Big[\sum_{j\in\mathcal{N}_i}\sum_{l\in\mathcal{N}_i\setminus\{j\}}\big\| \sigma_{G,j}'\cdot(\bm{w}_{i,ns}^{(k-1)}||\bm{w}_{j,ns}^{(k-1)})\\
					& - \sigma_{G,l}'\cdot(\bm{w}_{i,ns}^{(k-1)}||\bm{w}_{l,ns}^{(k-1)}) \big\|_2^2\Big].
				\end{aligned}
			\end{equation*}
			Then to satisfy Assumption \ref{a4}, the value of the fusion parameter $\mu_i^{(k)}$ in the $k$-th round for node $i\in\mathcal{V}$  should be constrained in
			\begin{equation}
				\label{e10}
				1-\frac{1}{\sqrt{d_i(d_i-1)D_i^{(k)}}}\le \mu_i^{(k)} \le 1.
			\end{equation} 
			
		\end{theorem}
		
		\newtheorem{remark}{Remark}
		\begin{remark}
			\label{r1}
			As the number of neighboring nodes $d_i$ becomes larger, the lower bound of $\mu_i^{(k)}$ increases.
			Moreover, the value of $D_i^{(k)}$ reflects the degree of non-i.i.d. of the neighboring nodes to some extent. If $x_j\ge 0$ is satisfied for all $j\in\mathcal{N}_i$, $D_i^{(k)}$ can be further simplified into the following expression:
			\begin{align}
			D_i^{(k)}=&\Big[\sum_{j\in\mathcal{N}_i}\big\|\bm{w}_{j,ns}^{(k-1)}\|_2^2\Big]\times\notag\\
   &\Big[\sum_{j\in\mathcal{N}_i}\sum_{l\in\mathcal{N}_i\setminus\{j\}} \|\bm{w}_{j,ns}^{(k-1)}-\bm{w}_{l,ns}^{(k-1)}\|_2^2\Big].
			\end{align}
			It can be observed that when the parameters of neighboring nodes are closer, which indicates that the non-i.i.d. degree is smaller, the lower bound of $\mu_i^{(k)}$ reduces. This is rational since a small value of $\mu_i^{(k)}$ represents more impact of the aggregation term under smaller non-i.i.d. degree. 
		\end{remark}
		
		Theorem \ref{t3} provides the lower bound of the fusion parameter, below which the convergence of GATTA cannot be guaranteed. We refer the readers to Appendix C for detailed proof.
		In the practical implementations, we do not focus on the fusion parameter design for each single node. For the sake of simplicity, we denote the fusion parameter for all the nodes by $\mu$ as applied in (\ref{fuse2}) and choose its value through experiments. 
		
		\section{Numerical Experiments}\label{experiment}
		In this section, we numerically evaluate the performance of our proposed algorithms under non-i.i.d. conditions. 
		In particular, we consider a multi-agent communication network with $N$ nodes, whose topology is generated randomly using the \emph{Erdos\_Renyi} random graph model, with the connectivity probability equal to $p$.
		{If not specified, we apply the widely-used AlexNet architecture in each agent, which is a representative DNN and CNN architecture.} The node-specific layer gets its parameters according to (\ref{fuse2}) or (\ref{fuse3}), including weights and biases. Meanwhile, to make the loss function smooth, we apply ELU as the activation functions for the whole network.
		
		The performance is evaluated on the image classification problem. 
		To validate the algorithm more comprehensively, we simulate on two different settings of non-i.i.d.: label distribution skew \cite{Li2021} and feature distribution skew \cite{Ghosh2020}. 
		For the label distribution skew, we consider the $10$-class classification problem over CIFAR-10 \cite{cifar} and randomly choose $c_i$ labels assigned for each agent, which reflects the non-i.i.d. data distribution. The training samples corresponding to the same label are averaged and randomly assigned to the agents. The testing samples are assigned to agents corresponding to their local label distributions. 
		Note that the number of training samples with respect to one label is averaged among agents with that label. Meanwhile, for a general and comprehensive evaluation of the learned model, the testing samples are assigned to agents with all of them corresponding to the local label distribution.
		
		For the feature distribution skew, we consider the $62$-class classification problem over FEMNIST \cite{Caldas2018}, which contains images of different characters written by different writers. We randomly assign $e_i$ different writers with their written characters for each agent and use $75\%$ of them for training, $25\%$ for testing. 
        The performance metric is the average of the testing accuracy among the agents.
		We compare the proposed GATTA and CE-GATTA with three baseline methods: centralized FL \cite{FL1}, D-SGD \cite{Lian2017} as well as independent learning in each agent (IL).
		For all distributed learning methods, agents exchange messages after one epoch of local training. 
		
		\subsection{Evaluation of convergence on the Different Datasets} \label{dif_dataset}
\begin{figure*}[!htp]
			\centering
			\includegraphics[width=0.8\textwidth]{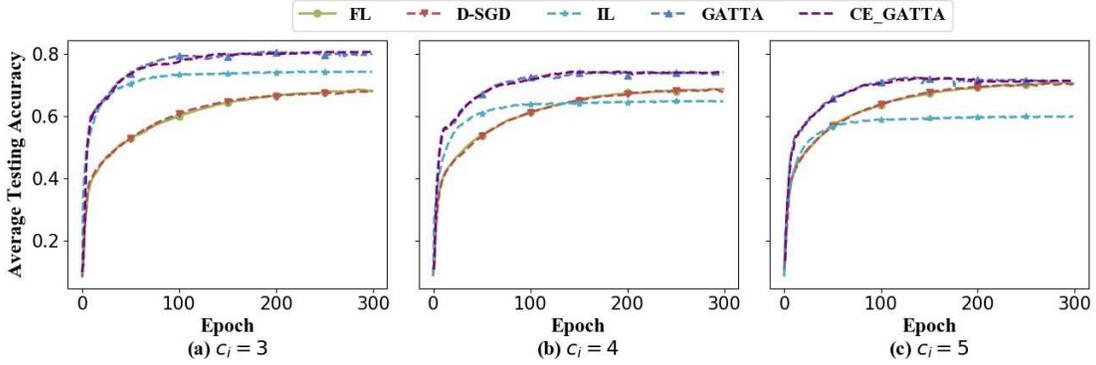}
			\caption{Convergence behaviours for CIFAR-10 under various $c_i$}
			\label{exp2}
		\end{figure*}

		\textbf{Label Distribution Skew.} 
		We first show the results on the CIFAR-10 dataset under different numbers of local labels $c_i=3,4,5$ and different numbers of local training samples $n_i$.
		The communication network is generated randomly with $N=100$ and $p=0.6$.
		In each agent, the local neural network is made of two $5\times5$ convolutional layers, each followed by a $3\times3$ max pooling layer with stride $2$, and three fully connected layers. The last fully-connected layer is the node-specific layer. For all the algorithms, the local optimizer is RMSProp \cite{RMSprop}.
		The learning rate of IL is set to $\eta=0.01$, while for others $\eta=0.001$. All of the learning rates are tuned from $\{0.1, 0.01, 0.001, 0.0001\}$ and we set $\mu=0.9$ through experiments.
		{The threshold for CE-GATTA is set to $\tau_i=\frac{1}{4d_i}$.}
		The results are shown in Figure \ref{exp2}. 
		
		As shown in Fig. \ref{exp2}, in the label skew condition, GATTA and CE-GATTA outperform the baseline methods in both convergence rate as well as resultant accuracy. 
		Here FL and D-SGD shows similar results due to the average and consensus procedure. Also, theorem in \cite{Lian2017} proves the same convergence rate of the centralized and decentralized method. 
		Secondly, comparing the results under different $c_i$, it can be observed that the superiority of GATTA over FL and D-SGD is more pronounced under smaller $c_i$. This is because a smaller $c_i$ indicates less relativity among agents and the personalization technique is more effective. Meanwhile, the independent learning method IL performs best in $c_i=3$. 
		
		Moreover, the communication-efficient implementation of GATTA shows almost the same convergence property as original GATTA. This is resulted from that the information transmission CE-GATTA removed is redundant or useless, and has little impact on the performance of the algorithm.
		In this way, the reduction of communication is effective without increasing the iteration number.
		
\begin{figure*}[!htp]
			\centering
			\vspace{-0.1cm}
			\includegraphics[width=0.8\textwidth]{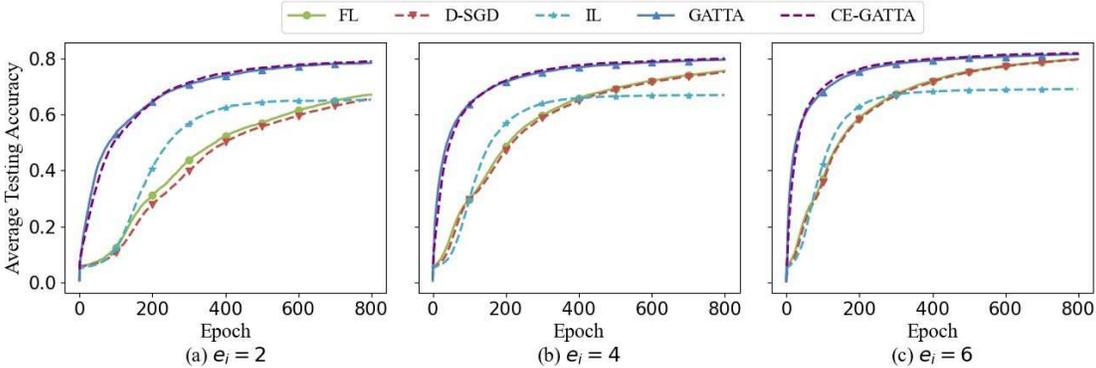}
			\caption{Convergence behaviours for FEMNIST under various $e_i$}
			\label{exp3}
                \vspace{-0.3cm}
		\end{figure*}

		\textbf{Feature Distribution Skew.}
		We next evaluate the performance in feature distribution skew condition through assigning different writers in FEMNIST for different agents. 
		The number of writers in each agent is set to $e_i=2,4,6$ respectively.
		The local network is made of two $3\times3$ convolutional layers, each followed by a $2\times2$ max pooling layer with stride $2$, and two fully connected layers, the last one of which serves as the node-specific layer.
		We set $\eta=0.01$ for all the algorithms, as tuned from $\{0.1, 0.01, 0.001, 0.0001\}$, and $\mu=0.7$, {$\tau_i=1/d_i$}. The results are shown in Figure \ref{exp3}.
		As shown in Fig. \ref{exp3}, in the feature skew condition, GATTA and CE-GATTA also show good performance compared with the other methods. It can be observed that the convergence rate of proposed algorithms is much faster than FL and D-SGD, indicating that the the proposed methods can quickly and effectively capture the useful information from other agents. Also, as $e_i$ reduces, indicating a larger degree of non-i.i.d., the GATTA shows significant accuracy performance compared with other baseline methods, highlighting the effectiveness of proposed algorithms in the non-i.i.d. conditions.
		
		{In both Fig. \ref{exp2} and \ref{exp3}, there exists a similar and inspiring trend that the superiority of GATTA and CE-GATTA, over FL and D-SGD, is higher under smaller number of labels or writers in each agent. To shed more light on its inherent reasons, we provide the following Remark \ref{r2}.
			
			\begin{remark}
				\label{r2}
				The local number of labels ($c_i$) or writers ($e_i$) affects the non-i.i.d. degree among nodes, which becomes higher when $c_i$ or $e_i$ decreases. When the non-i.i.d. degree becomes higher, the correlation among nodes reduces, leading to a worse performance of consensus learning methods such as FL or D-SGD and a higher superiority of personalized GATTA/CE-GATTA. On the contrary, when $c_i$ or $e_i$ increases, the correlation among nodes becomes larger, where a consensus model may adapt more on local data distribution and the superiority of GATTA/CE-GATTA becomes smaller. 
			\end{remark}
		}
		
		\subsection{Evaluation of accuracy On Different Network Topologies}
		We investigate the performance of the algorithms under different network topologies. Specifically, we use FEMNIST for validation and set $e_i=2$. The results under different numbers of agents as well as different probabilities of connectivity are evaluated. We set $\eta=0.01$ (except for the ring topology), $\mu=0.7$, $\tau_i=1/d_i$ and the maximum number of rounds is $800$.
		The results are first compared under fixed $N=100$ and different probabilities of connectivity $p=0.05, 0.2$. Then, we fix $p=0.2$ and set $N=50, 150$. 
		We also consider a more extreme condition of a ring communication network topology with $N=50$ nodes, where $\eta=0.008$.
		We additionally compare the algorithms with {four} different state-of-the-art methods as follows: 
		\begin{itemize}
			\item{The first one is the method in \cite{Collins2021} generalized in decentralized network, which we term as RepDL. In RepDL, each agent aggregate the parameters in the global component while updating the node-specific parameters only with local dataset. The comparison with RepDL can shows the effectiveness of the graph-based aggregation procedure. Its learning rate is $\eta=0.01$.}
			\item{The second procedure is the traditional D-SGD following fine-tuning on different nodes for personalization, termed as DSGD-FT. Such idea has achieved good performance in federated learning. Its learning rate is $\eta=0.01$.}
			\item{The third method is the $D^2$ training algorithm proposed in \cite{Tang2018}, whose learning rate is $\eta=0.1$.}
			\item{The last method is the GT-DSGD algorithm in \cite{Ran2021}, where the decaying step-size $\eta_k = \frac{1.0}{10+\sqrt{k}}$ are adopted and the Metropolis rule is applied to define the weight matrix as suggested by \cite{Pu2021}.
				$$
				\alpha_{ij} = \left\{
				\begin{aligned}
					&1/\max\{d_i, d_j\}  &\text{if} \quad j\in\mathcal{N}_i,\\
					&1-\sum\nolimits_{l\in\mathcal{N}_i}\alpha_{il}  &\text{if}\quad j=i,\\
					&0  &\text{otherwise}.\\
				\end{aligned}
				\right.
				$$}
		\end{itemize}
		The communication networks are generated randomly using the \emph{Erdos\_Renyi} model {and the testing accuracy results are averaged over $5$ trails as reported in Table \ref{NT_fe}, along with the $95\%$ confidence intervals.}
		As shown in Table \ref{NT_fe}, the proposed algorithms have the best performance among all algorithms, even under the sparse connectivity $p=0.05$ and extreme condition of ring communication network. 
		{Moreover, the resultant accuracy of CE-GATTA is similar to that of GATTA. 
		This is because of the mechanism of CE-GATTA, which can learn and adjust to fuse the information from the selected nodes. And such fusion may utilize the whole information flowing over the communication network.}
		$D^2$ algorithm fails to converge when $N=100,p=0.05$, $N=50,p=0.2$ and in the ring topology.
		
		
		\begin{table*}[!htp]
			\caption{Comparison of average testing accuracy under different network topologies.}
			\label{NT_fe}
			\centering
			\renewcommand\arraystretch{1.5}
			\begin{tabular}{llllll}
				\toprule
				\multicolumn{1}{c} {\multirow{2}{*}{Algorithms}} &  \multicolumn{5}{c}{Network Parameters} \\
				\cmidrule(r){2-6}
				$\quad$      & $N=100, p=0.05$  & $N=100, p=0.2$   & $N=50, p=0.2$   & $N=150, p=0.2$ & $N=50$, ring  \\
				\midrule
				{FL} & $67.39\pm0.13\%$ & $67.44\pm0.15\%$ & $67.20\pm 0.16\%$ & $67.40\pm 0.14\%$ & $67.15\pm 0.14\%$  \\
				{D-SGD} & $65.91\pm0.27\%$ &$66.08\pm0.16\%$  & $66.33\pm 0.25\%$   &   $66.21\pm 0.21\%$ & $66.26 \pm 0.20\%$  \\
				{IL} & $65.49\pm0.06\%$ & $65.53\pm0.05\%$ & $65.07\pm 0.09\%$  & $65.14\pm 0.05\%$ & $65.06\pm 0.10\%$ \\
				{RepDL} & $66.93\pm0.10\%$ & $66.85\pm0.13\%$ & $66.40\pm0.21\%$ & $66.54\pm0.19\%$ & $66.41\pm 0.06\%$\\
				{DSGD-FT} & $71.02\pm0.12\%$ & $72.46\pm0.10\%$  & $70.92\pm0.17\%$ &$70.95\pm0.15\%$& $72.24\pm 0.20\%$\\
				{$D^2$} & $-$ &$73.92\pm0.30\%$  & $-$ &$72.95\pm0.21\%$ & $-$\\
				{GT-DSGD} & $69.79\pm0.11\%$ &$72.57\pm0.16\%$  & $71.27\pm0.39\%$ &$72.34\pm0.35\%$ &$73.28\pm 0.21\%$\\
				\midrule
				{\textbf{GATTA}} & $\bm{78.78\pm0.21\%}$ &$\bm{78.81\pm0.23\%}$  & $\bm{78.90\pm0.15\%}$ &$\bm{79.03\pm0.20\%}$ & $\bm{77.10\pm 0.27\%}$\\
				{\textbf{CE-GATTA}} & $\bm{78.70\pm0.20\%}$ &$\bm{78.67\pm0.25\%}$  & $\bm{79.04\pm0.17\%}$ &$\bm{79.05\pm0.23\%}$ & $\bm{76.92\pm 0.28\%}$\\
				\bottomrule
			\end{tabular}
		\end{table*}
		
		{
			\subsection{Generalization to Other DNN Architecture}
			In this subsection, we simulate the proposed GATTA on other kinds of local neural network architecture.
			Different from the AlexNet above, we apply ResNet-18 \cite{ResNet} for CIFAR-10 and MLP for FEMNIST.
			Specifically, in ResNet-18, each convolutional layer is followed by a batch-normalization layer, whose shift and scale are trainable parameters.
			The MLP is a $784-400-100-62$ architecture with three fully connected (FC) layers, and each of the first two FC layers is followed by a batch-normalization layer.
			The learning rate is $0.001$ for all methods on ResNet-18, and $0.1$ for all approaches on MLP.
			In both networks, the last fully-connected layer is treated as the node-specific layer for GATTA.
			Moreover, we compare the results with another method proposed in \cite{Fedbn2021}, where the batch-normalization layers are not averaged in the training process and only trained with local data. We name it as BN-DSGD.
			The other settings are same as those in Section \ref{dif_dataset} and we set $c
			_i=3$ for CIFAR-10, $e_i=2$ for FEMNIST.
			The results in ResNet-18 and MLP are shown in Fig. \ref{ResNet} and \ref{MLP}.
			\begin{figure}[!htp]
				\centering
				\includegraphics[width=0.35\textwidth]{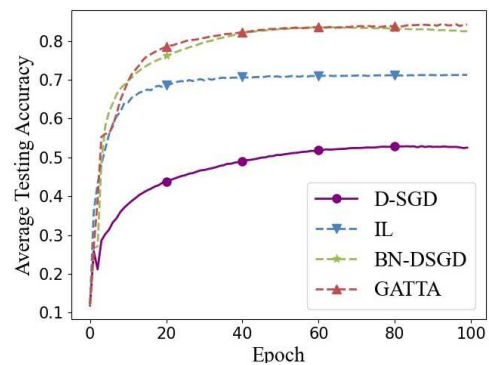}
				\caption{Convergence behaviors for CIFAR-10 on ResNet-18}
				\label{ResNet}
                \vspace{-0.2cm}
			\end{figure}
			
			\begin{figure}[!htp]
				\centering
				\includegraphics[width=0.35\textwidth]{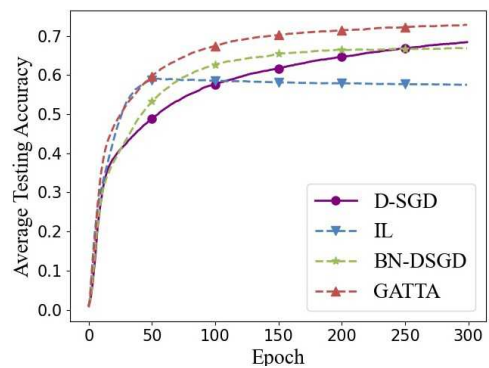}
				\caption{Convergence behaviors for FEMNIST on MLP}
				\label{MLP}
                \vspace{-0.2cm}
			\end{figure}
			
			In Fig. \ref{ResNet}, GATTA and BN-DSGD share similar performance, while in Fig. \ref{MLP}, GATTA outperforms BN-DSGD. 
			Note that BN-DSGD requires the network architecture having the batch-normalization layer and its performance highly relies on the number of local data samples training the batch-normalization layers. 
			Meanwhile, the results validate the efficiency and superiority of the proposed GATTA on different local DNN architectures.
		}
		
		\subsection{Evaluation of Communication Cost}
\begin{figure}[!htp]
			\centering
			\includegraphics[width=0.4\textwidth]{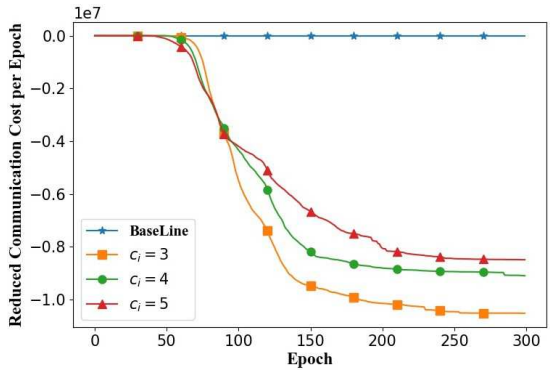}
			\caption{The reduced communication cost with epoch}
			\label{com_fig}
		\end{figure}

		In this part, we evaluate the communication cost of CE-GATTA and compare with the traditional D-SGD method. Note that here we focus on the decentralized communication network topology without a fusion center, so we do not conduct FL for comparison.
		Specifically, we measure the communication cost by the total number of parameters transmitted. 
		The algorithms stop when they achieve the accuracy requirements ($0.79\%, 0.75\%, 0.72\%$ for $c_i=3, 4, 5$ respectively) or the maximum iteration number.
		The setting of the simulation is the same as the label skew condition in Section \ref{dif_dataset}.
		We first show the reduction of communication cost with epoch in Fig. \ref{com_fig}, where the BaseLine refers to the methods of D-SGD or GATTA, which transmits all the parameters to all the neighboring nodes. 
		From Fig. \ref{com_fig}, it can be observed that as the iteration goes on, the communication cost of CE-GATTA per epoch reduces by stopping the transmission of less important parameters. Moreover, when the learning of the weight specification comes to converge, the condition of $c_i=3$ takes the least communication cost. It is rational since a smaller $c_i$ indicates less relativity among agents and there can be more ineffective information stopped to be transmitted.
		
		Then we show the results of total communication cost in Table. \ref{Com_cost}. It can be observed that compared with traditional D-SGD, CE-GATTA largely reduces the communication cost resulted from the faster convergence rate and less information transmission per epoch.
		
		\begin{table}[!htp]
			\caption{Comparison of communication cost}
			\label{Com_cost}
			\centering
			\begin{tabular}{llll}
				\toprule
				\quad & $c_i=3$ & $c_i=4$ &$c_i=5$\\
				\midrule
				D-SGD & $4.0204\times 10^{12}$ & $4.0204\times 10^{12}$ & $4.0204\times 10^{12}$\\
				\midrule
				CE-GATTA & $1.6076\times 10^{12}$ & $1.6578\times 10^{12}$ & $1.6580\times 10^{12}$\\
				\midrule
				Reduction & $60.0\%$ & $58.8\%$ & $58.8\%$ \\
				\bottomrule
			\end{tabular}
		\end{table}
		
		{
			In the following, we focus on the performance of CE-GATTA under different threshold $\tau_i$. 
			As we talked in Section \uppercase\expandafter{\romannumeral3}-C, when the threshold is small or the number of epochs is large, there exists little difference of the resultant accuracy over different $\tau_i$.
			To better show the difference and reveal the trade-off, we choose $\tau_i$ with relatively large values, where $\tau_i=1/d_i, 2/d_i, 3/d_i, 4/d_i$. And the total communication cost is calculated until the accuracy achieves $78\%$.
			Then we present the following Table \ref{Com_cost2} to show the comparison of communication cost.
		}
		
		\begin{table}[!htp]
			\caption{Comparison of communication cost ($\times 10^{12}$)}
			\label{Com_cost2}
			\centering
			\begin{tabular}{lllll}
				\toprule
				$\tau_i=1/4d_i$ & $\tau_i=1/d_i$ & $\tau_i=2/d_i$ & $\tau_i=3/d_i$ & $\tau_i=4/d_i$\\
				\midrule
				$1.0553$ & $1.0347$ & $1.2353$ & $1.2055$ & $1.2657$\\
				\bottomrule
			\end{tabular}
		\end{table}
		
		{It can be observed that the communication cost does not necessarily become smaller with the increasing $\tau_i$, due to a larger number of epochs to achieve the required accuracy. 
			Consequently, there exists a best choice of the threshold for CE-GATTA saving the communication cost most.}
		
		\section{Conclusion}\label{conclusion}
		We considered the statistical heterogeneous problem in the decentralized deep learning and proposed a graph-attention-based personalization method called GATTA. The GATTA enables each agent to adaptively utilize the information from neighboring agents.
		This can be implemented through learning specify weights for different neighboring agents in the training process, based on which we designed a communication-efficient GATTA.
		We also derived the theoretical convergence properties of GATTA and provided the range of the fusion parameter.
		Finally, we compared the performances of the proposed algorithms with other distributed learning algorithm under different datasets, non-i.i.d. settings, and network topologies. 
		The experiment results validated the superiority of the proposed algorithms over conventional schemes.
		
		The algorithm with rigorous theoretical guarantees provides a broad impact on improving the local learning quality for applications that deploy decentralized learning. 
		Although the local personalized model and the experiments are based on deep neural networks, the proposed graph-attention-based personalization technique could be generalized to other learning networks with proper design. 
		{Thus, one of our future researching topics is to generalize the personalized model into other neural networks.
			Another important issue is the theoretical convergence analysis of CE-GATTA, which could shed more light on its overall performance w.r.t. communication and computation costs.}
		Additionally, it is promising to apply the proposed algorithm to practical wireless communication problems, such as collaborative location for multiple base stations.
		
		\section*{Appendix}
		\subsection{Proof of the Lemmas}
		The proof of Lemma \ref{l1} is as follows.
		\begin{proof}
			\begin{align}
				& \mathbb{E}[\|\Delta_{i}^{(k)}\|_2^2] = \mathbb{E}[\|\sum_{t=0}^{T}g_{i,t}^{(k)}\|_2^2] \notag\\
				& \overset{(a)}{=} \mathbb{E}[\|\sum_{t=0}^{T-1}(g_{i,t}^{(k)}-\nabla f_i(\bm{w}_{i,t}^{(k)}))\|_2^2]+ \mathbb{E}[\|\sum_{t=0}^{T-1}\nabla f_i(\bm{w}_{i,t}^{(k)})\|_2^2]\notag\\
				& \overset{(b)}{\le} T\cdot \chi^2 + \mathbb{E}[\|\sum_{t=0}^{T-1}\nabla f_i(\bm{w}_{i,t}^{(k)})\|_2^2],
			\end{align}
			where (a) follows from the fact that $\mathbb{E}[\|\bm{x}\|_2^2]=\mathbb{E}[\|\bm{x}-\mathbb{E}[x]\|_2^2]+\|\mathbb{E}[x]\|_2^2$ and (b) follows from the unbiased estimator.
		\end{proof}
		
		The proof of Lemma \ref{l2} is as follows, which is similar to that of Lemma 2 in \cite{Haibo2021}.
		\begin{proof}
			We have that 
			\begin{align}
				&\mathbb E[\|(\bm{w}_{i,t}^{(k)}-\bm{w}_{i}^{(k)})\|_2^2] = \mathbb E[\|(\bm{w}_{i,t-1}^{(k)}-\bm{w}_{i}^{(k)})-\eta g_{i,t-1}^{(k)}\|_2^2]\notag\\
				&= \mathbb E[\|(\bm{w}_{i,t-1}^{(k)}-\bm{w}_{i}^{(k)})-\eta (g_{i,t-1}^{(k)}-\nabla f_i(\bm{w}_{i,t-1}^{(k)})\notag\\
				&\quad +\nabla f_i(\bm{w}_{i,t-1}^{(k)})-\nabla f_i(\bm{w}_{i}^{(k)})+\nabla f_i(\bm{w}_{i}^{(k)}))\|_2^2]\notag\\
				&\overset{(a)}{\le} \mathbb E[\|(\bm{w}_{i,t-1}^{(k)}-\bm{w}_{i}^{(k)})-\eta (\nabla f_i(\bm{w}_{i,t-1}^{(k)})-\nabla f_i(\bm{w}_{i}^{(k)})\notag\\
				&\quad+\nabla f_i(\bm{w}_{i}^{(k)}))\|_2^2]+E[\|\eta (g_{i,t-1}^{(k)}-\nabla f_i(\bm{w}_{i,t-1}^{(k)})\|_2^2]\notag\\
				&\overset{(b)}{\le} (1+\frac{1}{2T-1})\mathbb E[\|\bm{w}_{i,t-1}^{(k)}-\bm{w}_{i}^{(k)}\|_2^2]\notag\\
				&\quad+(1+2T-1)\mathbb E[\|\eta (\nabla f_i(\bm{w}_{i,t-1}^{(k)})-\nabla f_i(\bm{w}_{i}^{(k)})\notag\\
				&\quad+\nabla f_i(\bm{w}_{i}^{(k)}))\|_2^2]+E[\|\eta (g_{i,t-1}^{(k)}-\nabla f_i(\bm{w}_{i,t-1}^{(k)})\|_2^2]\notag\\
				& \le (1+\frac{1}{2T-1})\mathbb E[\|\bm{w}_{i,t-1}^{(k)}-\bm{w}_{i}^{(k)}\|_2^2]\\
				&\quad+4T\mathbb E[\|\eta (\nabla f_i(\bm{w}_{i,t-1}^{(k)})-\nabla f_i(\bm{w}_{i}^{(k)}))\|_2^2]\notag\\
				& \quad+4T\mathbb E[\|\eta\nabla f_i(\bm{w}_{i}^{(k)}))\|_2^2]+E[\|\eta (g_{i,t-1}^{(k)}-\nabla f_i(\bm{w}_{i,t-1}^{(k)})\|_2^2]\notag\\
				&\le (1+\frac{1}{2T-1}+4T\eta^2L^2)\mathbb E[\|\bm{w}_{i,t-1}^{(k)}-\bm{w}_{i}^{(k)}\|_2^2]\notag \\
				& \quad+4T\mathbb E[\|\eta\nabla f_i(\bm{w}_{i}^{(k)}))\|_2^2]+\eta^2\chi^2\notag \\
				&\overset{(c)}{\le} (1+\frac{1}{T-1})\mathbb E[\|\bm{w}_{i,t-1}^{(k)}-\bm{w}_{i}^{(k)}\|_2^2]\notag \\
				&\quad +4T\mathbb E[\|\eta\nabla f_i(\bm{w}_{i}^{(k)}))\|_2^2]+\eta^2\chi^2,\notag
			\end{align}
			where (a) follows from Assumption \ref{a2} that $g_{i,t-1}^{(k)}$ is an unbiased estimation of $\nabla f_i(\bm{w}_{i,t-1}^{(k)})$. (b) follows from $(x+y)^2\le(1+\frac{1}{K})x^2+(1+K)y^2$ and (c) follows from $\eta<\frac{1}{24TL}$.
			
			Unrolling the recursion, we get
			\begin{align}
				& \mathbb E[\|(\bm{w}_{i,t}^{(k)}-\bm{w}_{i}^{(k)})\|_2^2] \le \\
				& \quad\sum_{p=0}^{t-1}(1+\frac{1}{T-1})^p\Big[\eta^2\chi^2+4T\mathbb E[\|\eta\nabla f_i(\bm{w}_{i}^{(k)}))\|_2^2]\Big]\notag\\
				& \le (T-1)\times[(1+\frac{1}{T-1})^T-1]\times\Big[\eta^2\chi^2+4T\mathbb E[\|\eta\nabla f_i(\bm{w}_{i}^{(k)}))\|_2^2]\Big]\notag\\
				& \le 4T\eta^2\chi^2+16T^2\eta^2\mathbb E[\|\nabla f_i(\bm{w}_{i}^{(k)}))\|_2^2],\notag
			\end{align}
			where the last inequality follows from $(1+\frac{1}{T-1})^T\le5$ for $T>1$.
		\end{proof}
		
		The proof is Lemma \ref{l3} is as follows.
		\begin{proof}
			Provided that $\eta\le\frac{1}{24TL}$, we have
			\begin{align}
				\label{pl2_1}
				&-\mathbb{E}[\eta\nabla f_{i}(\bm{w}_{i}^{(k)})^T\Delta_i^{(k)}] = -\eta\mathbb E<\nabla f_{i}(\bm{w}_{i}^{(k)}), {\sum_{t=0}^{T-1}}g_{i,t}^{(k)}>\notag\\
				& = -\eta\mathbb E<\nabla f_{i}(\bm{w}_{i}^{(k)}), {\sum_{t=0}^{T-1}}\nabla f_i(\bm{w}_{i,t}^{(k)})>\notag\\
				& \overset{(a)}{\le} -\frac{T\eta}{2}\|\nabla f_{i}(\bm{w}_{i}^{(k)})\|_2^2- \frac{\eta}{2T}\mathbb E[\|\sum_{t=0}^{T-1}\nabla f_i(\bm{w}_{i,t}^{(k)})\|_2^2]\notag\\
				&\quad+\frac{\eta}{2}\sum_{t=0}^{T-1}\mathbb E[\|(\nabla f_i(\bm{w}_{i,t}^{(k)})-\nabla f_{i}(\bm{w}_{i}^{(k)}))\|_2^2]\notag\\
				& \overset{(b)}{\le} -\frac{T\eta}{2}\|\nabla f_{i}(\bm{w}_{i}^{(k)})\|_2^2- \frac{\eta}{2T}\mathbb E[\|\sum_{t=0}^{T-1}\nabla f_i(\bm{w}_{i,t}^{(k)})\|_2^2]\notag\\
				&\quad +\frac{L^2\eta}{2}\sum_{t=0}^{T-1}\mathbb E[\|(\bm{w}_{i,t}^{(k)}-\bm{w}_{i}^{(k)})\|_2^2]\notag\\
				& \overset{(c)}{\le} -\frac{T\eta}{2}\|\nabla f_{i}(\bm{w}_{i}^{(k)})\|_2^2-\frac{\eta}{2T}\mathbb E[\|\sum_{t=0}^{T-1}\nabla f_i(\bm{w}_{i,t}^{(k)})\|_2^2]\notag\\
				&\quad+\frac{L^2\eta T}{2}[4T\eta^2\chi^2+16T^2\eta^2\|\nabla f_{i}(\bm{w}_{i}^{(k)})\|_2^2]\notag\\
				& = -T\eta(\frac{1}{2}-8\eta^2T^2L^2)\|\nabla f_{i}(\bm{w}_{i}^{(k)})\|_2^2+2\eta^3T^2L^2 \chi^2\notag\\
				&\quad- \frac{\eta}{2T}\mathbb E[\|\sum_{t=0}^{T-1}\nabla f_i(\bm{w}_{i,t}^{(k)})\|_2^2],
			\end{align}
			where (a) follows from $\|\sum_{i=1}^n\bm{a}_i\|_2^2\le n\sum_{i=1}^n\|\bm{a}_i\|_2^2$, (b) follows from Assumption \ref{a1} and (c) follows from Lemma \ref{l2}.
			Then according to Lemma \ref{l1}, we have 
			\begin{align}
				& \mathbb{E}\Big[-\eta\nabla f_{i}(\bm{w}_{i}^{(k)})^T\Delta_i+\frac{1}{2}\eta^2L\|\Delta_i^{(k)}\|^2_2 \Big] \label{26}\\ 
				& \le -T\eta(\frac{1}{2}-8\eta^2T^2L^2)\|\nabla f_{i}(\bm{w}_{i}^{(k)})\|_2^2+2\eta^3T^2L^2\chi^2 \notag\\
				&\quad+ \frac{1}{2}\eta^2LT\chi^2+(\frac{\eta^2L}{2}-\frac{\eta}{2T})\mathbb E[\|\sum_{t=0}^{T-1}\nabla f_i(\bm{w}_{i,t}^{(k)})\|_2^2] \notag\\
				& \overset{(a)}{\le}  -T\eta(\frac{1}{2}-8\eta^2T^2L^2)\|\nabla f_{i}(\bm{w}_{i}^{(k)})\|_2^2+\frac{\eta^2TL}{2}(1+4\eta TL)\chi^2, \notag
			\end{align}
			where (a) follows from $\eta<\frac{1}{24TL}<\frac{1}{TL}$.
		\end{proof}
    The proof of Lemma \ref{l4} is as follows.
    \begin{proof}
        \begin{align}
	&\mathbb{E}\Big\|\frac{1}{N}\sum_{i\in\mathcal{V}}\nabla f_i(\bar{\bm{w}}_{g}^{(k)})-\frac{1}{N}\sum_{i\in\mathcal{V}}\nabla f_i(\bm{w}_{i,g}^{(k)})\Big\|^2_2 \notag\\
	& = \frac{1}{N^2} \mathbb{E}\Big\|\sum_{i\in\mathcal{V}}\big[\nabla f_i(\bar{\bm{w}}_g^{(k)})- \nabla f_i(\bm{w}_{i,g}^{(k)})\big]\Big\|^2_2\notag\\
 &\le \frac{L^2}{N}\sum_{i\in\mathcal{V}}\mathbb{E}\|\bar{\bm{w}}_g^{(k)}-\bm{w}_{i,g}^{(k)}\|_2^2, \notag
    \end{align}
 where the last inequality follows from the $L$-smoothness of the local functions.
    \end{proof}
The proof of Lemma \ref{l5} is as follows.
\begin{proof}
\begin{align}
    & \bar{\bm{w}}_g^{(k)}-\bar{\bm{w}}_g^{(k-1)}= \frac{1}{N}\textbf{W}_{g}^{(k)}\textbf{1}_N-\frac{1}{N}(\textbf{W}_{g}^{(k-1)})\textbf{1}_N\notag \\
    & \quad= \frac{1}{N}(\textbf{W}_{g}^{(k-1)}-\eta\Xi_{g}^{(k-1)})\tilde{\textbf{A}}\textbf{1}_N-\frac{1}{N}(\textbf{W}_{g}^{(k-1)})\textbf{1}_N\notag \\
    &\quad = -\frac{\eta}{N}\Xi_{g}\tilde{\textbf{A}}\textbf{1}_N= -\frac{\eta}{N}\sum_{i\in\mathcal{V}}\Delta_{i,g}^{(k-1)}.
\end{align}
where the last equality follows from Assumption \ref{a6} that $\tilde{\textbf{A}}$ is a symmetric doubly stochastic.
\end{proof}

		\subsection{Proof of Theorem \ref{t1}}
		\begin{proof}
			The proof is inspired by the analysis in SGD method as in \cite{Bottou2018} and D-SGD method as in \cite{Li2019}.
			We treat the parameters $\bm{w}_{i,lu},\bm{\beta}_{i}$ and $\bar{\bm{w}}_{g}$ respectively so as to correspond to their update rules.
			
			Without loss of generality, we consider one specific node $i$ in the following proof. 
			According to the smoothness of the local objective function, we have
			\begin{align}
				\label{pt1_1}
				&f_i(\tilde{\bm{v}}_i^{(k)})- f_i(\tilde{\bm{v}}_i^{(k-1)}) \le \\
				& \nabla f_i(\tilde{\bm{v}}_i^{(k-1)})^T(\tilde{\bm{v}}_i^{(k)}-\tilde{\bm{v}}_i^{(k-1)})+\frac{L}{2}\|\tilde{\bm{v}}_i^{(k)}-\tilde{\bm{v}}_i^{(k-1)}\|^2_2\notag\\
				&\overset{(a)} = \nabla f_i(\bm{w}_{i,lu}^{(k-1)})^T(\bm{w}_{i, lu}^{(k)}-\bm{w}_{i, lu}^{(k-1)})+\frac{L}{2}\|\bm{w}_{i, lu}^{(k)}-\bm{w}_{i, lu}^{(k-1)}\|^2_2\notag\\
				& \quad + \nabla f_i(\bm{\beta}_{i}^{(k-1)})^T(\bm{\beta}_{i}^{(k)}-\bm{\beta}_{i}^{(k-1)})+\frac{L}{2}\|\bm{\beta}_{i}^{(k)}-\bm{\beta}_{i}^{(k-1)}\|^2_2\notag\\
				&\quad + \nabla f_i(\bar{\bm{w}}_{g}^{(k-1)})^T(\bar{\bm{w}}_{g}^{(k)}-\bar{\bm{w}}_{g}^{(k-1)})+\frac{L}{2}\|\bar{\bm{w}}_{g}^{(k)}-\bar{\bm{w}}_{g}^{(k-1)}\|^2_2, \notag
			\end{align}
			where (a) follows from the definition that $\tilde{\bm{v}}_i$ is the concatenation of the averaged $\bar{\bm{w}_{g}}$ and individual $\bm{w}_{i,lu}$, $\bm{\beta}_i$.
			According to (\ref{e2}) and (\ref{e3}), the first four terms in the right side of (\ref{pt1_1}) is equal to
			\begin{equation}
				\label{pt1_2}
				\begin{aligned}
					& -\eta\nabla f_i(\bm{w}_{i,lu}^{(k-1)})^T\Delta_{i,lu}^{(k-1)}+\frac{1}{2}\eta^2L\|\Delta_{i,lu}^{(k-1)}\|^2_2 \\
					& -\eta \nabla f_i(\bm{\beta}_{i}^{(k-1)})^T\Delta_{i,b}^{(k-1)}+\frac{1}{2}\eta^2L\|\Delta_{i,b}^{(k-1)}\|^2_2.
				\end{aligned}
			\end{equation}
			Take the expectation of (\ref{pt1_1}) on the both sides, we have
			\begin{align}
				\label{pt1_4}
				&\mathbb{E}\big[f_i(\tilde{\bm{v}}_i^{(k)})- f_i(\tilde{\bm{v}}_i^{(k-1)})\big]  \le \\
				& \mathbb{E}\Big[-\eta\nabla f_i(\bm{w}_{i,lu}^{(k-1)})^T\Delta_{i,lu}^{(k-1)}-\eta\nabla f_i(\bm{\beta}_{i}^{(k-1)})^T\Delta_{i,b}^{(k-1)}\notag\\
				&\quad +\frac{1}{2}\eta^2L\|\Delta_{i,lu}^{(k-1)}\|^2_2 + \frac{1}{2}\eta^2L\|\Delta_{i,b}^{(k-1)}\|^2_2\Big] \notag\\
				&\quad + \mathbb{E}\Big[  \nabla f_i(\bar{\bm{w}}_{g}^{(k-1)})^T(\bar{\bm{w}}_{g}^{(k)}-\bar{\bm{w}}_{g}^{(k-1)})\Big] \notag\\
				& \quad+\frac{L}{2}\mathbb{E}\Big[\|\bar{\bm{w}}_{g}^{(k)}-\bar{\bm{w}}_{g}^{(k-1)}\|^2_2\Big]. \notag
			\end{align}
			We now focus on the right side of (\ref{pt1_4}). Under Lemma \ref{l3}, the first expectation term can be bounded by
			\begin{align}
				\label{pt1_5}
				& \mathbb{E}\Big[-\eta\nabla f_i(\bm{w}_{i,lu}^{(k-1)})^T\Delta_{i,lu}^{(k-1)}-\eta\nabla f_i(\bm{\beta}_{i}^{(k-1)})^T\Delta_{i,b}^{(k-1)} \notag\\
				& \quad +\frac{1}{2}\eta^2L\|\Delta_{i,lu}^{(k-1)}\|^2_2 + \frac{1}{2}\eta^2L\|\Delta_{i,b}^{(k-1)}\|^2_2\Big] \\
				& \le -cT\eta( \|\nabla f_{i}(\bm{w}_{i,lu}^{(k-1)})\|_2^2+\|\nabla f_{i}(\bm{\beta}_{i}^{(k-1)})\|_2^2)\notag \\
				&\quad +\eta^2TL(1+4\eta TL)\chi^2.\notag
			\end{align}
			
	We add both sides of (\ref{pt1_4}) from $i=1$ to $i=N$, and derive the results by N, then we have 
    \begin{align}
        \label{pt1_11}
        &\frac{1}{N}\sum_{i\in\mathcal{V}}\mathbb{E}\big[f_i(\tilde{\bm{v}}_i^{(k)})- f_i(\tilde{\bm{v}}_i^{(k-1)})\big] \notag \\
        &\le \frac{-cT\eta}{N}\sum_{i\in\mathcal{V}}( \|\nabla f_{i}(\bm{w}_{i,lu}^{(k-1)})\|_2^2+\|\nabla f_{i}(\bm{\beta}_{i}^{(k-1)})\|_2^2)\notag \\
        & \quad+ \eta^2TL(1+4\eta TL)\chi^2 \notag\\
        & \quad+\mathbb{E}\big[\frac{1}{N}\sum_{i\in\mathcal{V}}\nabla f_i(\bar{\bm{w}}_{g}^{(k-1)})^T(\bar{\bm{w}}_{g}^{(k)}-\bar{\bm{w}}_{g}^{(k-1)})\big] \notag\\
        & \quad+\frac{L}{2}\mathbb{E}\Big[\|\bar{\bm{w}}_{g}^{(k)}-\bar{\bm{w}}_{g}^{(k-1)}\|^2_2\Big].
    \end{align}
 
        When Assumption \ref{a1}-\ref{a3} and \ref{a4} hold, then the last two terms in (\ref{pt1_11}) can be bounded by 
			\begin{align}
				\label{pt1_6}
				&\mathbb{E}\big[\frac{1}{N}\sum_{i\in\mathcal{V}}\nabla f_i(\bar{\bm{w}}_{g}^{(k-1)})^T(\bar{\bm{w}}_{g}^{(k)}-\bar{\bm{w}}_{g}^{(k-1)})\big] \notag\\
        & \quad +\frac{L}{2}\mathbb{E}\Big[\|\bar{\bm{w}}_{g}^{(k)}-\bar{\bm{w}}_{g}^{(k-1)}\|^2_2\Big]\notag\\
		& \overset{(a)}{=} -\eta\mathbb{E}\Big[\nabla F(\bar{\bm{w}}_{g}^{(k-1)})^T\bar{\Delta}_{g}^{(k-1)}\Big] + \frac{L\eta^2}{2}\mathbb{E}\|\bar{\Delta}_{g}^{(k-1)}\|_2^2 \notag \\
  & = -\eta\mathbb{E}\Big[\nabla F(\bar{\bm{w}}_{g}^{(k-1)})^T(\bar{\Delta}_{g}^{(k-1)}-\sum_{t=0}^{T-1}\nabla F(\bm{w}_{g,t}^{(k-1)}))\Big] \notag \\
  &\quad -\eta \mathbb{E}\Big[\nabla F(\bar{\bm{w}}_{g}^{(k-1)})^T\sum_{t=0}^{T-1}\nabla F(\bm{w}_{g,t}^{(k-1)})\Big] \notag \\
  &\quad +\frac{L\eta^2}{2}\mathbb{E}\|\bar{\Delta}_{g}^{(k-1)}-\sum_{t=0}^{T-1}\nabla F(\bm{w}_{g,t}^{(k-1)})+\sum_{t=0}^{T-1}\nabla F(\bm{w}_{g,t}^{(k-1)})\|_2^2 \notag \\
  & \overset{(b)}{\le}-\frac{\eta T}{2} \mathbb{E}\|\nabla F(\bar{\bm{w}}_{g}^{(k-1)})\|_2^2 -\frac{\eta T}{2}(\frac{1}{6}-LT\eta)\mathbb{E}\|\nabla F(\bm{w}_g^{(k-1)})\|_2^2 \notag \\
  & \quad + \frac{4\eta TL^2}{3N}\sum_{i\in\mathcal{V}}\mathbb{E}\|\bar{\bm{w}}_g^{(k-1)}-\bm{w}_{i,g}^{(k-1)}\|_2^2\notag \\
  & \quad +\frac{1}{N}\big[T\eta^2L(4\kappa^2T+\chi^2) + 6T^2\eta^3\chi^2L^2\big], 
\end{align}
where (a) follows from the definition $\nabla F(\bar{\bm{w}}_g)\triangleq\frac{1}{N}\sum_{i\in\mathcal{V}}\nabla f_i(\bar{\bm{w}}_{g})$ and Lemma \ref{l5}. The derivation of (b) deals with $\sum_{t=0}^{T-1}\nabla F(\bm{w}_{g,t}^{(k-1)})$ by subtracting and then adding $T\nabla F(\bm{w}_g^{(k-1)})$. It also requires $\eta<\frac{1}{24TL}$. Its detailed derivation is omitted here.
		
Define $\bar{\bar{\textbf{A}}}_{s, k-1} = \prod_{l=s}^{k-1}\tilde{\textbf{A}}$, $\textbf{Q}=\frac{1}{N}\textbf{1}_N\textbf{1}_N^T$ and $\rho_{s, k-1}=\|\bar{\bar{\textbf{A}}}_{s, k-1}-\textbf{Q}\|$.
Then based on Lemma 2-6 in \cite{Li2019}, when the learning rate is sufficiently small such that $\eta<\frac{1}{24TL}$ and $\eta<\frac{1}{32TL\sqrt{C_K}}$, following some derivations under multiple steps of SGD and the adjustment of Lemma 2-6 in \cite{Li2019}, we could also derive that
\begin{align}
\label{pt1_8}
&\frac{1}{N}\sum_{i\in\mathcal{V}}\sum_{k=1}^{K}\mathbb{E}\|\bar{\bm{w}}_g^{(k-1)}-\bm{w}_{i,g}^{(k-1)}\|_2^2\notag\\
\le & \Big[A_K\chi^2+B_KT(\kappa^2+T\eta^2\chi^2L^2)+ \\
&\quad \frac{C_KT}{K}\sum_{k=1}^{K}\mathbb{E}\big\|\nabla F(\bm{w}_{g}^{(k-1)})\big\|_2^2\Big]\times \frac{24\eta^2TK}{1-48\eta^2L^2T^2C_K}, \notag
\end{align}
where $A_K, B_K, C_K$ are defined as follows.
\begin{align}
&A_K=\frac{1}{K}\sum_{k=1}^{K}\sum_{s=1}^{k-1}\rho_{s, k-1}^2,\quad B_K=\frac{1}{K}\sum_{k=1}^{K}\Big(\sum_{s=1}^{k-1}\rho_{s, k-1}\Big)^2,\notag \\
& C_K = \max_{s\in[K-1]}\sum_{k=s+1}^{K}\rho_{s, k-1}\Big(\sum_{l=1}^{k-1}\rho_{l, k-1}\Big),\notag
\end{align}
			
Recall the definition of the global objective function as (\ref{opt_sys}), we have $F(\tilde{\bm{V}}^{(k)})=\frac{1}{N}\sum_{i\in\mathcal{V}}f_i(\tilde{\bm{v}}_i^{(k)})$ and the left side of (\ref{pt1_11}) is equal to $F(\tilde{\bm{V}}^{(k)})-F(\tilde{\bm{V}}^{(0)})$.
Then we add both the right side and left side of (\ref{pt1_11}) from $k=1$ to $K$, we could derive the following expression.
\begin{align}
\label{pt1_9}
&\mathbb{E} [F(\tilde{\bm{V}}^{(K)})-F(\tilde{\bm{V}}^{(0)})] \notag\\
& \le -\frac{cT\eta}{N}\sum_{i\in\mathcal{V}}\sum_{k=0}^{K-1}\|\nabla f_i(\bm{v}_{i,ns}^{(k)})\|_2^2 -{cT\eta}\sum_{k=0}^{K-1}\|\nabla F(\bar{\bm{w}}_{g}^{(k)})\|_2^2\notag \\
&\quad -\frac{\eta T}{2}(\frac{1}{6}-LT\eta-128\eta^2T^2L^2C_K)\sum_{k=0}^{K-1}\mathbb{E}\|\nabla F(\bm{w}_g^{(k)})\|_2^2 \notag\\
&\quad +\eta^2TLK(1+4\eta TL)\chi^2 \notag\\
&\quad + \frac{1}{N}\big[T\eta^2LK(4\kappa^2T+\chi^2) + 6T^2\eta^3\chi^2L^2K\big] \notag\\
  & \quad + 64\eta^3T^2L^2K(A_K\chi^2+B_KT( \kappa^2+T\eta^2\chi^2L^2)).
\end{align}
requiring the learning rate satisfies $1-48\eta^2L^2T^2C_K\ge \frac{1}{2}$. 
If we constrain $\eta<\frac{1}{24TL}$ and $\eta<\frac{1}{32TL\sqrt{C_K}}$, we have 
$$
\frac{1}{6}-LT\eta-128\eta^2T^2L^2C_K>0,
$$
and we can derive that
			\begin{align}
				\label{pt1_10}
				&\mathbb{E}\big[F(\tilde{\bm{V}}^{(K)})- F(\tilde{\bm{V}}^{(0)})\big] \\
    &\le-\frac{cT\eta}{N}\sum_{i\in\mathcal{V}}\sum_{k=0}^{K-1}\big[\|\nabla f_i(\bm{v}_{i,ns}^{(k)})\|_2^2+\|\nabla F(\bar{\bm{w}}_{g}^{(k)})\|_2^2\big]+E. \notag
			\end{align}
			Here 
			\begin{align}
			& E= \eta^2TLK(1+4\eta TL)\chi^2 \notag\\
&\quad + \frac{1}{N}\big[T\eta^2LK(4\kappa^2T+\chi^2) + 6T^2\eta^3\chi^2L^2K\big] \notag\\
  & \quad + 64\eta^3T^2L^2K(A_K\chi^2+B_KT( \kappa^2+T\eta^2\chi^2L^2)).\notag
			\end{align}
			
   We also have
    \begin{align}
    -\frac{1}{N}\sum_{i\in\mathcal{V}}\|\nabla f_i(\bm{v}_{i,ns}^{(k)})\|_2^2 &\le -\big\|\frac{1}{N}\sum_{i\in\mathcal{V}}\nabla f_i(\bm{v}_{i,ns}^{(k)})\big\|_2^2 \notag\\
    & = -\big\|\nabla F(\bm{v}_{ns}^{(k)}) \big\|_2^2. \notag
    \end{align}
   
   Rearrange the terms in (\ref{pt1_10}), we can derive 
			\begin{align}
				\label{pt4_1}
				&\frac{1}{K}\sum\nolimits_{k=0}^{K-1}\times\mathbb{E}\big[\|\nabla F(\bm{v}_{ns}^{(k)})\|_2^2+\|\nabla F(\bar{\bm{w}}_{g}^{(k)})\|_2^2\big] \notag\\ 
				& \le\frac{\mathbb{E}\big[F(\tilde{\bm{V}}^{(0)})- F(\tilde{\bm{V}}^{(K)})\big]}{cTK\eta} + \frac{E}{cTK\eta}, \notag
			\end{align}
			which implies that 
			\begin{equation}
				\min_{k\in[K]}\mathbb{E}\big[\|\nabla F(\bm{v}_{ns}^{(k)})\|_2^2+\|\nabla F(\bar{\bm{w}}_{g}^{(k)})\|_2^2\big]\le \frac{F_0-F_*}{cTK\eta}+\Phi,
			\end{equation}
			where
			\begin{align}
			& \Phi= \frac{1}{c}\Big\{\eta L(1+4\eta TL)\chi^2 \notag\\
&\quad + \frac{1}{N}\big[\eta L(4\kappa^2T+\chi^2) + 6T\eta^2\chi^2L^2\big] \notag\\
  & \quad + 64\eta^2TL^2(A_K\chi^2+B_KT( \kappa^2+T\eta^2\chi^2L^2))\Big\}.\notag
			\end{align}
			 This completes the proof.
		\end{proof}
		
		\subsection{Proof of Theorem \ref{t3}}
		\begin{proof}
			In Theorem \ref{t3}, we derive the appropriate range of the fusion parameter $\mu$ so as to sufficiently satisfy Assumption \ref{a4}. We first derive the expression of $\nabla f_i(\bm{w}_{i, lu})$ and the expression of $\nabla f_i(\bm{\beta}_{i})$.
			According to the aggregation model (\ref{fuse2}) and the gradient back propagation, we can derive that $\nabla f_i(\bm{w}_{i, lu})=\mu\nabla f_i(\bm{w}_{i, ns})$.
			Then we have
			\begin{equation}
				\|\nabla f_i(\bm{w}_{i, lu})\|_2^2=\mu^2\times\|\nabla f_i(\bm{w}_{i, ns})\|_2^2. 
			\end{equation}
			According to Assumption \ref{a4}, we have $\|\nabla f_i(\bm{w}_{i, lu})\|_2^2\le G$ and $\|\nabla f_i(\bm{w}_{i, ns})\|_2^2\le G$. Then it can be derived that $0\le\mu\le1$.
			
			Following we derive the lower bound of $\mu$ in the $k$-th round for node $i$, which we denote by $\mu_{i}^{(k)}$ for better clarification.
			For simplicity, we denote the gradient value of $\sigma_G({x}_j)$ by $\sigma_{G,j}'$, where ${x}_j\triangleq {\bm{\beta}_i^{(k-1)}}^T(\bm{w}_{i,ns}^{(k-1)}||\bm{w}_{j,ns}^{(k-1)})$.
			And we denote $f_e(j)=\exp(\sigma_{G,  j})$ for node $i$. Likewise, the gradient value of $\sigma$ is denoted by $\sigma'$.
			In this paper, $\sigma_G$ is the ELU activation function.
			Then it can be derived that the gradient's value at the $\bm{\beta}_i^{(k-1)}$ is:
			\begin{align}
				\label{pt3_2}
				& \nabla f_i(\bm{\beta}_{i}^{(k-1)})=\nabla f_i(\bm{w}_{i, ns}^{(k-1)})\times  \frac{(1-\mu_i^{(k)})\sigma'^T}{{\big[\sum_{j\in\mathcal{N}_i}f_e(j)\big]}^2}\times\notag\\
				&\quad\sum_{j\in\mathcal{N}_i}\sum_{l\in\mathcal{N}_i\setminus\{j\}}\Big\{\bm{w}_{j,ns}^{(k-1)}\times f_e(j)\times f_e(l)\times\\
				&\big[\sigma_{G,j}'^T\cdot(\bm{w}_{i,ns}^{(k-1)}||\bm{w}_{j,ns}^{(k-1)}) - \sigma_{G,l}'^T\cdot(\bm{w}_{i,ns}^{(k-1)}||\bm{w}_{l,ns}^{(k-1)}) \big]\Big\}.\notag
			\end{align}
			The detailed derivation is omitted here for simplicity. Take the $\ell_2$ norm on both sides of (\ref{pt3_2}) and it can be derived that
			\begin{align}
				\label{pt3_3}
				& \|\nabla f_i(\bm{\beta}_{i}^{(k-1)})\|_2^2=  (1-\mu_i^{(k)})^2\|\nabla f_i(\bm{w}_{i, ns}^{(k-1)})\|_2^2\times \\ 
				& \quad \frac{\|\sigma'\|_2^2}{{\big[\sum_{j\in\mathcal{N}_i}f_e(j)\big]}^4}\Big\|\sum_{j\in\mathcal{N}_i}\sum_{l\in\mathcal{N}_i\setminus\{j\}}\Big\{\bm{w}_{j,ns}^{(k-1)}\times f_e(j)\times f_e(l)\times\notag\\
				& \quad \big[\sigma_{G,j}'\cdot(\bm{w}_{i,ns}^{(k-1)}||\bm{w}_{j,ns}^{(k-1)}) - \sigma_{G,l}'\cdot(\bm{w}_{i,ns}^{(k-1)}||\bm{w}_{l,ns}^{(k-1)}) \big]\Big\}\Big\|_2^2.\notag
			\end{align}
			Denote 
			\begin{align}
				& f_G \triangleq \Big\|\sum_{j\in\mathcal{N}_i}\sum_{l\in\mathcal{N}_i\setminus\{j\}}\Big\{\bm{w}_{j,ns}^{(k-1)} f_e(j) f_e(l)\notag\\ 
				&\quad\times\big[\sigma_{G,j}'\cdot(\bm{w}_{i,ns}^{(k-1)}||\bm{w}_{j,ns}^{(k-1)}) - \sigma_{G,l}'\cdot(\bm{w}_{i,ns}^{(k-1)}||\bm{w}_{l,ns}^{(k-1)}) \big]\Big\}\Big\|_2^2 \notag \\
				& =\Big\|\sum_{j\in\mathcal{N}_i}\bm{w}_{j,ns}^{(k-1)}\sum_{l\in\mathcal{N}_i\setminus\{j\}}\Big\{ f_e(j) f_e(l)\times\notag\\
				&\quad \big[\sigma_{G,j}'\cdot(\bm{w}_{i,ns}^{(k-1)}||\bm{w}_{j,ns}^{(k-1)}) - \sigma_{G,l}'\cdot(\bm{w}_{i,ns}^{(k-1)}||\bm{w}_{l,ns}^{(k-1)}) \big]\Big\}\Big\|_2^2 \notag \\
				& \overset{(a)}\le d_i\sum_{j\in\mathcal{N}_i}\Big\|\bm{w}_{j,ns}^{(k-1)}\sum_{l\in\mathcal{N}_i\setminus\{j\}}f_e(j) f_e(l)\times\notag\\
				&\quad \big[\sigma_{G,j}'\cdot(\bm{w}_{i,ns}^{(k-1)}||\bm{w}_{j,ns}^{(k-1)}) - \sigma_{G,l}'\cdot(\bm{w}_{i,ns}^{(k-1)}||\bm{w}_{l,ns}^{(k-1)}) \big]\Big\|_2^2 \notag\\
				& \overset{(b)}\le d_i\Big[\sum_{j\in\mathcal{N}_i}\big\|\bm{w}_{j,ns}^{(k-1)}\|_2^2\Big] \times\Big[\sum_{j\in\mathcal{N}_i}\Big\| \sum_{l\in\mathcal{N}_i\setminus\{j\}}f_e(j) f_e(l)\times \notag\\
				&\quad \big[\sigma_{G,j}'\cdot(\bm{w}_{i,ns}^{(k-1)}||\bm{w}_{j,ns}^{(k-1)}) - \sigma_{G,l}'\cdot(\bm{w}_{i,ns}^{(k-1)}||\bm{w}_{l,ns}^{(k-1)}) \big] \Big\|_2^2\Big] \notag  \\
				& \overset{(c)}\le d_i(d_i-1)\Big[\sum_{j\in\mathcal{N}_i}\big\|\bm{w}_{j,ns}^{(k-1)}\|_2^2\Big]\times \Big[\sum_{j\in\mathcal{N}_i}\sum_{l\in\mathcal{N}_i\setminus\{j\}}\Big\| f_e(j) f_e(l)\notag\\
				&\quad \times\big[\sigma_{G,j}'\cdot(\bm{w}_{i,ns}^{(k-1)}||\bm{w}_{j,ns}^{(k-1)}) - \sigma_{G,l}'\cdot(\bm{w}_{i,ns}^{(k-1)}||\bm{w}_{l,ns}^{(k-1)}) \big] \Big\|_2^2\Big] \notag\\
				&\overset{(d)}\le d_i(d_i-1)\Big[\sum_{j\in\mathcal{N}_i}\big\|\bm{w}_{j,ns}^{(k-1)}\|_2^2\Big]\times\notag\\
				&\quad\Big[\sum_{j\in\mathcal{N}_i}\sum_{l\in\mathcal{N}_i\setminus\{j\}}\Big\| f_e(j) f_e(l)\Big\|_2^2\Big]\notag\\
				&\quad \times\Big[\sum_{j\in\mathcal{N}_i}\sum_{l\in\mathcal{N}_i\setminus\{j\}}\Big\| \sigma_{G,j}'\cdot(\bm{w}_{i,ns}^{(k-1)}||\bm{w}_{j,ns}^{(k-1)}) -\notag\\
				&\quad \quad \sigma_{G,l}'\cdot(\bm{w}_{i,ns}^{(k-1)}||\bm{w}_{l,ns}^{(k-1)}) \Big\|_2^2\Big] \notag,
			\end{align}
			where (a) and (c) follows from $\|\sum_{i=1}^n\bm{a}_i\|_2^2\le n\sum_{i=1}^n\|\bm{a}_i\|_2^2$, (b) and (d) follows from $\sum_i\|\bm{a}_i\|_2^2\|\bm{b}_i\|_2^2\le\sum_i\|\bm{a}_i\|_2^2\sum_i\|\bm{b}_i\|_2^2$.
			
			With $\|\nabla f_i(\bm{w}_{i, ns})\|_2^2\le G$ according to Assumption \ref{a4}, $\|\sigma'\|_2^2\le1$ according to (\ref{a4_2}) in Assumption \ref{a4} and 
			\begin{equation}
				\label{pt3_6}
				0\le\Big[\frac{\sum_{j\in\mathcal{N}_i}\sum_{l\in\mathcal{N}_i\setminus\{j\}}[f_e(j) f_e(l)]^2}{(\sum_{m\in\mathcal{N}_i}f_e(m))^4 }\Big]\le 1, 
			\end{equation}
			then we have that 
			\begin{align}
				\label{pt3_5}
				& \|\nabla f_i(\bm{\beta}_{i}^{(k-1)})\|_2^2\le (1-\mu_i^{(k)})^2G
				d_i(d_i-1)\Big[\sum_{j\in\mathcal{N}_i}\big\|\bm{w}_{j,ns}^{(k-1)}\|_2^2\Big]\times \notag\\
				& \quad \Big[\sum_{j\in\mathcal{N}_i}\sum_{l\in\mathcal{N}_i\setminus\{j\}}\Big\| \sigma_{G,j}'\cdot(\bm{w}_{i,ns}^{(k-1)}||\bm{w}_{j,ns}^{(k-1)}) -\notag\\
				&\quad \quad \sigma_{G,l}'\cdot(\bm{w}_{i,ns}^{(k-1)}||\bm{w}_{l,ns}^{(k-1)}) \Big\|_2^2\Big].
			\end{align}
			We denote the last two terms by
			\begin{align}
				D_i^{(k)}\triangleq & \Big[\sum_{j\in\mathcal{N}_i}\big\|\bm{w}_{j,ns}^{(k-1)}\|_2^2\Big]\times \notag\\
				& \Big[\sum_{j\in\mathcal{N}_i}\sum_{l\in\mathcal{N}_i\setminus\{j\}}\Big\| \sigma_{G,j}'\cdot(\bm{w}_{i,ns}^{(k-1)}||\bm{w}_{j,ns}^{(k-1)}) -\\
				& \quad \sigma_{G,l}'\cdot(\bm{w}_{i,ns}^{(k-1)}||\bm{w}_{l,ns}^{(k-1)}) \Big\|_2^2\Big].\notag
			\end{align}
			
			Then it sufficiently satisfies Assumption \ref{a4} if $(1-\mu_i^{(k)})^2Gd_i(d_i-1)D_i^{(k)}\le G$, following which we can derive the lower bound of $\mu$ for node $i$ in the $k$-th round as 
			\begin{equation}
				\label{pt3_8}
				\mu_i^{(k)}\ge 1-\frac{1}{\sqrt{d_i(d_i-1)D_i^{(k)}}}.
			\end{equation}
		\end{proof}

  \newpage
  \section*{Supplementary Details}
  In this section, we provide some detailed derivations in the proof of Theorem 1 in Appendix B. Specifically, the detailed derivations of (32) and (33) are shown as follows.
    
    \subsection*{A. Definitions}
		For simplicity, we make the following definitions, where the first two are defined the same as those in the manuscript and the last one is only used in the proof.
		\begin{align}
			& \nabla F(\bar{\bm{w}}_g^{(k)})\triangleq\frac{1}{N}\sum_{i\in\mathcal{V}}\nabla f_i(\bar{\bm{w}}_{g}^{(k)}) \notag \\
			& \nabla F(\bm{w}_g^{(k)})\triangleq\frac{1}{N}\sum_{i\in\mathcal{V}}\nabla f_i(\bm{w}_{i,g}^{(k)}) \notag\\
			& \nabla \bm F(\textbf{W}_g)\triangleq [\nabla f_1(\bm{w}_{1,g}), ..., \nabla f_N(\bm{w}_{N,g})].
		\end{align}
		
		Note that under the definition of $\nabla \bm F(\textbf{W}_g)$, we have $\|\nabla \bm F(\textbf{W}_g^{k})\|_F^2 = \sum_{i\in\mathcal{V}}\|\nabla f_i(\bm{w}_{i,g}^{(k)}))\|_2^2$.
		\subsection*{B. Useful Lemmas and their proofs}	
		
		\begin{lemma}
			\label{l6}
			\begin{align}
				\label{pl5_1}
				& \mathbb{E}\|\nabla F(\bm{w}_g^{(k-1)})-\frac{1}{T}\sum_{t=0}^{T-1}\nabla F(\bm{w}_{g,t}^{(k-1)})\|_2^2 \notag \\
				& \le  \frac{4T\eta^2\chi^2L^2}{N}+\frac{16T^2\eta^2L^2}{N^2}\|\nabla \bm F(\textbf{W}_g^{(k-1)})\|_F^2 .
			\end{align}
		\end{lemma}
		The proof of Lemma \ref{l6} is as follows.
		\begin{proof}
			\begin{align}
				\label{pl5_2}
				& \mathbb{E}\|\nabla F(\bm{w}_g^{(k-1)})-\frac{1}{T}\sum_{t=0}^{T-1}\nabla F(\bm{w}_{g,t}^{(k-1)})\|_2^2 \notag \\
				& = \frac{1}{N^2}\mathbb{E}\|\frac{1}{T}\sum_{i\in\mathcal{V}}\sum_{t=0}^{T-1}\nabla f_i(\bm{w}_{i,g}^{(k-1)})-\frac{1}{T}\sum_{i\in\mathcal{V}}\sum_{t=0}^{T-1}\nabla f_i(\bm{w}_{i,g,t}^{(k-1)})\|_2^2 \notag \\
				& \overset{(a)}\le \frac{1}{N^2T^2}\times T\sum_{i\in\mathcal{V}}\sum_{t=0}^{T-1}\mathbb{E}\|\nabla f_i(\bm{w}_{i,g}^{(k-1)})-\nabla f_i(\bm{w}_{i,g,t}^{(k-1)})\|_2^2\notag \\
				& \overset{(b)}\le \frac{L^2}{N^2T}\sum_{i\in\mathcal{V}}\sum_{t=0}^{T-1}\mathbb{E}\|\bm{w}_{i,g}^{(k-1)}-\bm{w}_{i,g,t}^{(k-1)}\|_2^2\notag\\
				& \overset{(c)}\le \frac{L^2}{N^2T}\sum_{i\in\mathcal{V}}\sum_{t=0}^{T-1}(4T\eta^2\chi^2+16T^2\eta^2\|\nabla f_i(\bm{w}_{i,g}^{(k-1)}))\|_2^2 \notag \\
				& \overset{(d)}= \frac{4T\eta^2\chi^2L^2}{N}+\frac{16T^2\eta^2L^2}{N^2}\|\nabla \bm F(\textbf{W}_g^{(k-1)})\|_F^2 ,
			\end{align}
			where (a) follows from that each node independently work in the $(k-1)$-th round before communication, (b) follows from the $L$-smoothness of the local function, (c) follows from Lemma 2, and (d) follows from the definition of $\nabla \bm F(\textbf{W})$.
		\end{proof}

		\begin{lemma}
			\label{l7}
			\begin{align}
				& -\mathbb{E}\|\sum_{t=0}^{T-1}\nabla F(\bm{w}_{g,t}^{(k-1)})\|_2^2 \notag\\
				& \le -\frac{T^2}{2}\mathbb{E}\|\nabla F(\bm{w}_g^{(k-1)})\|_2^2\notag\\
				&+T^2L^2\Big(\frac{4T\eta^2\chi^2}{N}+\frac{16T^2\eta^2}{N^2}\|\nabla \bm F(\textbf{W}_g^{(k-1)})\|_F^2\Big).
			\end{align}
		\end{lemma}
		The proof of Lemma \ref{l7} is as follows.
		\begin{proof}
			\begin{align}
				&\mathbb{E}\|\nabla F(\bm{w}_g^{(k-1)})\|_2^2 \notag\\
				&= \mathbb{E}\|\nabla F(\bm{w}_g^{(k-1)})-\frac{1}{T}\sum_{t=0}^{T-1}\nabla F(\bm{w}_{g,t}^{(k-1)})+\frac{1}{T}\sum_{t=0}^{T-1}\nabla F(\bm{w}_{g,t}^{(k-1)})\|_2^2 \notag\\
				&\le 2\mathbb{E}\|\nabla F(\bm{w}_g^{(k-1)})-\frac{1}{T}\sum_{t=0}^{T-1}\nabla F(\bm{w}_{g,t}^{(k-1)})\|_2^2\notag\\
				&\quad \frac{2}{T^2}\mathbb{E}\|\sum_{t=0}^{T-1}\nabla F(\bm{w}_{g,t}^{(k-1)})\|_2^2\notag\\
				& \overset{(a)}\le 2L^2\Big( \frac{4T\eta^2\chi^2}{N}+\frac{16T^2\eta^2}{N^2}\|\nabla \bm F(\textbf{W}_g^{(k-1)})\|_F^2\Big)\notag \\
				& \quad + \frac{2}{T^2}\mathbb{E}\|\sum_{t=0}^{T-1}\nabla F(\bm{w}_{g,t}^{(k-1)})\|_2^2,
			\end{align}
			where (a) follows from Lemma \ref{l6} This completes the proof.
		\end{proof}
		
		\begin{lemma}
			\label{l8}
			\begin{align}
				& \mathbb{E}\|\sum_{t=0}^{T-1}\nabla F(\bm{w}_{g,t}^{(k-1)})\|_2^2 \notag\\
				& \le 2T^2\mathbb{E}\|\nabla F(\bm{w}_g^{(k-1)})\|_2^2 \\
				& \quad + 2T^2L^2\Big(\frac{4T\eta^2\chi^2}{N}+\frac{16T^2\eta^2}{N^2}\|\nabla \bm F(\textbf{W}_g^{(k-1)})\|_F^2\Big).\notag
			\end{align}
		\end{lemma}
		The proof of Lemma \ref{l8} is as follows.
		\begin{proof}
			\begin{align}
				&\mathbb{E}\|\sum_{t=0}^{T-1}\nabla F(\bm{w}_{g,t}^{(k-1)})\|_2^2 \notag\\
				&= \mathbb{E}\|\sum_{t=0}^{T-1}\nabla F(\bm{w}_{g,t}^{(k-1)})-T\nabla F(\bm{w}_g^{(k-1)})+T\nabla F(\bm{w}_g^{(k-1)})\|_2^2 \notag\\
				&\le 2T^2\mathbb{E}\|\nabla F(\bm{w}_g^{(k-1)})-\frac{1}{T}\sum_{t=0}^{T-1}\nabla F(\bm{w}_{g,t}^{(k-1)})\|_2^2\notag\\
				&\quad +2T^2\mathbb{E}\|\nabla F(\bm{w}_g^{(k-1)})\|_2^2\notag\\
				& \overset{(a)}\le 2T^2L^2\Big( \frac{4T\eta^2\chi^2}{N}+\frac{16T^2\eta^2}{N^2}\|\nabla \bm F(\textbf{W}_g^{(k-1)})\|_F^2\Big)\notag \\
				& \quad + 2T^2\mathbb{E}\|\nabla F(\bm{w}_g^{(k-1)})\|_2^2,
			\end{align}
			where (a) follows from Lemma \ref{l6}. This completes the proof. 
		\end{proof}
		
		\begin{lemma}
			\label{l13}
			If $\eta<\frac{1}{24TL}$, then we have
			\begin{align}
				\mathbb{E}\|\sum_{t=0}^{T-1}\nabla \bm F(\textbf{W}_{g,t}^{(k)})\|_F^2  \le  & 3T^2\mathbb{E}\|\nabla \bm F(\textbf{W}_{g}^{(k)})\|_F^2 +8T^3\eta^2\chi^2L^2N. \notag
			\end{align}
		\end{lemma}
		The proof of Lemma \ref{l13} is as follows.
		\begin{proof}
			\begin{align}
				&\mathbb{E}\|\sum_{t=0}^{T-1}\nabla \bm F(\textbf{W}_{g,t}^{(k)})\|_F^2 \notag\\
				&= \mathbb{E}\|\sum_{t=0}^{T-1}\nabla \bm F(\textbf{W}_{g,t}^{(k)})-T{\nabla F}(\textbf{W}_g^{(k)})+T{\nabla F}(\textbf{W}_g^{(k)})\|_F^2 \notag\\
				&\le 2\sum_{i\in\mathcal{V}}\mathbb{E}\|\sum_{t=0}^{T-1}({\nabla f_i}(\bm{w}_{i,g,t}^{(k)})-{\nabla f_i}(\bm{w}_{i,g}^{(k}))\|_2^2\notag\\
				&\quad +2T^2\mathbb{E}\|{\nabla F}(\textbf{W}_g^{(k)})\|_F^2\notag\\
				& \overset{(a)}\le 2TL^2\sum_{i\in\mathcal{V}}\sum_{t=0}^{T-1}\mathbb{E}\|\bm{w}_{i,g,t}^{(k)}-\bm{w}_{i,g}^{(k)}\|_2^2\notag\\
				&\quad +2T^2\mathbb{E}\|{\nabla F}(\textbf{W}_g^{(k)})\|_F^2 \notag \\
				& \overset{(b)}\le 2T^2\mathbb{E}\|\nabla \bm F(\textbf{W}_{g}^{(k)})\|_F^2 +8T^3\eta^2\chi^2L^2N\notag\\
				&\quad +32T^4\eta^2L^2\sum_{i\in\mathcal{V}}\mathbb{E}\|\nabla f_i(\bm{w}_{i,g}^{(k)})\|_2^2, \notag\\
				& \overset{(c)}\le3T^2\mathbb{E}\|\nabla \bm F(\textbf{W}_{g}^{(k)})\|_F^2 +8T^3\eta^2\chi^2L^2N, 
			\end{align}
			where (a) follows from the $L$-smoothness of the local function, (b) follows from Lemma 2, and (c) follows from $\eta<\frac{1}{24TL}<\frac{1}{\sqrt{32}TL}$. This completes the proof.
		\end{proof}
		
		\begin{lemma}
			\label{l11}
			(Bound on second moments of gradients) Under Assumption 2 and 5, we have
			\begin{align}
				\frac{1}{N}\mathbb{E}\|\nabla \bm F(\textbf{W}_g^{(k)})\|_F^2\le & \frac{8L^2}{N}\sum_{i\in\mathcal{V}}\mathbb{E}\|\bm{w}_{i,g}^{(k)}-\bar{\bm{w}}_g^{(k)}\|_2^2 \notag \\
				& +4\kappa^2+4\mathbb{E}\|\nabla F(\bm{w}_g^{(k)})\|_2^2.\notag
			\end{align}
		\end{lemma}
		Lemma \ref{l11} is the same as Lemma 4 in [30], which is not related to the multiple steps' updating process and thus remains the same.
		
		\subsection*{C. Derivation of (32)}
		\begin{proof} 
			\begin{align}
				\label{pd32_1}
				&\mathbb{E}\Big[\nabla F(\bar{\bm{w}}_{g}^{(k-1)})^T(\bar{\bm{w}}_{g}^{(k)}-\bar{\bm{w}}_{g}^{(k-1)})\Big] +\frac{L}{2}\mathbb{E}\Big[\|\bar{\bm{w}}_{g}^{(k)}-\bar{\bm{w}}_{g}^{(k-1)}\|^2_2\Big]\notag\\
				& \overset{(a)}{=} -\eta\mathbb{E}\Big[\nabla F(\bar{\bm{w}}_{g}^{(k-1)})^T\bar{\Delta}_{g}^{(k-1)}\Big] + \frac{L\eta^2}{2}\mathbb{E}\|\bar{\Delta}_{g}^{(k-1)}\|_2^2 \notag \\
				& = -\eta\mathbb{E}\Big[\nabla F(\bar{\bm{w}}_{g}^{(k-1)})^T(\bar{\Delta}_{g}^{(k-1)}-\sum_{t=0}^{T-1}\nabla F(\bm{w}_{g,t}^{(k-1)}))\Big] \notag \\
				&\quad -\eta \mathbb{E}\Big[\nabla F(\bar{\bm{w}}_{g}^{(k-1)})^T\sum_{t=0}^{T-1}\nabla F(\bm{w}_{g,t}^{(k-1)})\Big] \notag \\
				&\quad +\frac{L\eta^2}{2}\mathbb{E}\|\bar{\Delta}_{g}^{(k-1)}-\sum_{t=0}^{T-1}\nabla F(\bm{w}_{g,t}^{(k-1)})+\sum_{t=0}^{T-1}\nabla F(\bm{w}_{g,t}^{(k-1)})\|_2^2 \notag \\
				& \overset{(b)}{=} -\eta \mathbb{E}\Big[\nabla F(\bar{\bm{w}}_{g}^{(k-1)})^T\sum_{t=0}^{T-1}\nabla F(\bm{w}_{g,t}^{(k-1)})\Big] \notag \\
				&\quad +\frac{L\eta^2}{2}\mathbb{E}\|\bar{\Delta}_{g}^{(k-1)}-\sum_{t=0}^{T-1}\nabla F(\bm{w}_{g,t}^{(k-1)})+\sum_{t=0}^{T-1}\nabla F(\bm{w}_{g,t}^{(k-1)})\|_2^2,
			\end{align}
			where (a) follows from Lemma 5. (b) follows from that $\Delta_{i,g} = \sum_{t=1}^{T-1}\bm{g}_{i,t}$ and $\mathbb{E}[\bm{g}_{i,t}]=\nabla f_i(\bm{w}_{i,t})$ according to Assumption 3.
			
			Then the first term in (\ref{pd32_1}) can be bounded by
			\begin{align}
				\label{pd32_2}
				& -\eta \mathbb{E}\Big<\nabla F(\bar{\bm{w}}_{g}^{(k-1)}), \sum_{t=0}^{T-1}\nabla F(\bm{w}_{g,t}^{(k-1)})\Big> \notag\\
				& \overset{(a)}= -\frac{\eta T}{2} \mathbb{E}\|\nabla F(\bar{\bm{w}}_{g}^{(k-1)})\|_2^2 -\frac{\eta T}{2}\mathbb{E}\|\frac{1}{T}\sum_{t=0}^{T-1}\nabla F(\bm{w}_{g,t}^{(k-1)})\|_2^2 \notag\\
				&\quad +\frac{\eta T}{2}\mathbb{E}\|\nabla F(\bar{\bm{w}}_{g}^{(k-1)})-\frac{1}{T}\sum_{t=0}^{T-1}\nabla F(\bm{w}_{g,t}^{(k-1)})\|_2^2 \notag\\
				& = -\frac{\eta T}{2} \mathbb{E}\|\nabla F(\bar{\bm{w}}_{g}^{(k-1)})\|_2^2 -\frac{\eta T}{2}\mathbb{E}\|\frac{1}{T}\sum_{t=0}^{T-1}\nabla F(\bm{w}_{g,t}^{(k-1)})\|_2^2 \notag \\
				&\quad +\frac{\eta T}{2}\mathbb{E}\|\nabla F(\bar{\bm{w}}_{g}^{(k-1)})-\nabla F(\bm{w}_g^{(k-1)})+\nabla F(\bm{w}_g^{(k-1)}) \notag\\
				& \quad\quad-\frac{1}{T}\sum_{t=0}^{T-1}\nabla F(\bm{w}_{g,t}^{(k-1)})\|_2^2 \notag\\
				& \overset{(b)}\le -\frac{\eta T}{2} \mathbb{E}\|\nabla F(\bar{\bm{w}}_{g}^{(k-1)})\|_2^2- \frac{\eta T}{2}\mathbb{E}\|\frac{1}{T}\sum_{t=0}^{T-1}\nabla F(\bm{w}_{g,t}^{(k-1)})\|_2^2 \notag \\
				&\quad +{\eta T}\mathbb{E}\|\nabla F(\bar{\bm{w}}_{g}^{(k-1)})-\nabla F(\bm{w}_g^{(k-1)})\|_2^2\notag\\
				& \quad +{\eta T}\mathbb{E}\|\nabla F(\bm{w}_g^{(k-1)})-\frac{1}{T}\sum_{t=0}^{T-1}\nabla F(\bm{w}_{g,t}^{(k-1)})\|_2^2\notag \\
				& \overset{(c)}\le -\frac{\eta T}{2} \mathbb{E}\|\nabla F(\bar{\bm{w}}_{g}^{(k-1)})\|_2^2-\frac{\eta T}{2}\mathbb{E}\|\frac{1}{T}\sum_{t=0}^{T-1}\nabla F(\bm{w}_{g,t}^{(k-1)})\|_2^2 \notag \\
				&\quad +{\eta T}\mathbb{E}\|\nabla F(\bar{\bm{w}}_{g}^{(k-1)})-\nabla F(\bm{w}_g^{(k-1)})\|_2^2\notag\\
				& \quad +{\eta TL^2}\Big(\frac{4T\eta^2\chi^2}{N}+\frac{16T^2\eta^2}{N^2}\|\nabla \bm F(\textbf{W}_{g}^{(k-1)})\|_F^2\Big),
			\end{align}
			where (a) follows from $<\bm{a}, \bm{b}>=\frac{1}{2}\|\bm{a}\|^2+\frac{1}{2}\|\bm{b}\|^2-\frac{1}{2}\|\bm{a}-\bm{b}\|^2$, (b) follows from $\|\bm{a}+\bm{b}\|_2^2\le 2(\|\bm{a}\|_2^2+\|\bm{b}\|_2^2)$ and (c) follows from Lemma \ref{l6}.
			
			The second term in (\ref{pd32_1}) can be bounded as follows.
			\begin{align}
				\label{pd32_3}
				&\frac{L\eta^2}{2}\mathbb{E}\|\bar{\Delta}_{g}^{(k-1)}-\sum_{t=0}^{T-1}\nabla F(\bm{w}_{g,t}^{(k-1)})+\sum_{t=0}^{T-1}\nabla F(\bm{w}_{g,t}^{(k-1)})\|_2^2 \notag \\
				& \le L\eta^2\mathbb{E}\|\bar{\Delta}_{g}^{(k-1)}-\sum_{t=0}^{T-1}\nabla F(\bm{w}_{g,t}^{(k-1)})\|_2^2 \notag\\
				&\quad + L\eta^2\mathbb{E}\|\sum_{t=0}^{T-1}\nabla F(\bm{w}_{g,t}^{(k-1)})\|_2^2 \notag \\
				& = L\eta^2\mathbb{E}\|\sum_{t=0}^{T-1}\bar{\bm{g}}_{g,t}^{(k-1)}-\sum_{t=0}^{T-1}\nabla F(\bm{w}_{g,t}^{(k-1)})\|_2^2\notag\\
				& \quad + L\eta^2\mathbb{E}\|\sum_{t=0}^{T-1}\nabla F(\bm{w}_{g,t}^{(k-1)})\|_2^2\notag\\
				& \overset{(a)}= L\eta^2\sum_{t=0}^{T-1} \mathbb{E}\Big\|\frac{1}{N}\sum_{i\in\mathcal{V}}\big[{\bm{g}}_{i,g,t}^{(k-1)}-\nabla f_i(\bm{w}_{i,g,t}^{(k-1)})\big]\Big\|_2^2\notag\\
				&\quad + L\eta^2\mathbb{E}\|\sum_{t=0}^{T-1}\nabla F(\bm{w}_{g,t}^{(k-1)})\|_2^2 \notag\\
				&\overset{(b)}\le \frac{L\eta^2\chi^2T}{N} + L\eta^2\mathbb{E}\|\sum_{t=0}^{T-1}\nabla F(\bm{w}_{g,t}^{(k-1)})\|_2^2,
			\end{align}
			where (a) follows from that $\mathbb{E}[\bar{\bm{g}}_{g,t}^{(k-1)}]=\nabla F(\bm{w}_{g,t}^{(k-1)})$ and (b) follows from Assumption 4 and the fact that in the $(k-1)$-th round, each node work independently with local dataset.
			
			Take the summation of (\ref{pd32_2}) and (\ref{pd32_3}), we can derive the upper bound of (\ref{pd32_1}) as follows.
			\begin{align}
				\label{pd32_4}
				&\mathbb{E}\Big[\nabla F(\bar{\bm{w}}_{g}^{(k-1)})^T(\bar{\bm{w}}_{g}^{(k)}-\bar{\bm{w}}_{g}^{(k-1)})\Big] +\frac{L}{2}\mathbb{E}\Big[\|\bar{\bm{w}}_{g}^{(k)}-\bar{\bm{w}}_{g}^{(k-1)}\|^2_2\Big]\notag\\
				& \le -\frac{\eta T}{2} \mathbb{E}\|\nabla F(\bar{\bm{w}}_{g}^{(k-1)})\|_2^2 - \frac{\eta T}{2}\mathbb{E}\|\frac{1}{T}\sum_{t=0}^{T-1}\nabla F(\bm{w}_{g,t}^{(k-1)})\|_2^2 \notag \\
				&\quad +{\eta T}\mathbb{E}\|\nabla F(\bar{\bm{w}}_{g}^{(k-1)})-\nabla F(\bm{w}_g^{(k-1)})\|_2^2\notag\\
				& \quad +{\eta TL^2}\Big(\frac{4T\eta^2\chi^2}{N}+\frac{16T^2\eta^2}{N^2}\|\nabla \bm F(\textbf{W}_g^{(k-1)})\|_F^2\Big)\notag\\
				& \quad+\frac{L\eta^2\chi^2T}{N}+ L\eta^2\mathbb{E}\|\sum_{t=0}^{T-1}\nabla F(\bm{w}_{g,t}^{(k-1)})\|_2^2\notag \\
				& = -\frac{\eta T}{2} \mathbb{E}\|\nabla F(\bar{\bm{w}}_{g}^{(k-1)})\|_2^2  + \frac{1}{N}(4T^2\eta^3\chi^2L^2 +LT\eta^2\chi^2)\notag\\
				&\quad -\eta(\frac{1}{2T}-L\eta)\mathbb{E}\|\sum_{t=0}^{T-1}\nabla F(\bm{w}_{g,t}^{(k-1)})\|_2^2\notag\\
				&\quad + \frac{16T^3\eta^3L^2}{N^2}\|\nabla \bm F(\textbf{W}_g^{(k-1)})\|_F^2\notag\\
				& \quad + {\eta T}\mathbb{E}\|\nabla F(\bar{\bm{w}}_{g}^{(k-1)})-\nabla F(\bm{w}_g^{(k-1)})\|_2^2.
			\end{align}
			Under the assumption that $\eta\le\frac{1}{24TL}$, it can be derived that $\frac{1}{2T}-L\eta\ge0$. Then based on Lemma \ref{l7}, (\ref{pd32_4}) can be further upper bounded by
			\begin{align}
				\label{pd32_5}
				&\mathbb{E}\Big[\nabla F(\bar{\bm{w}}_{g}^{(k-1)})^T(\bar{\bm{w}}_{g}^{(k)}-\bar{\bm{w}}_{g}^{(k-1)})\Big] +\frac{L}{2}\mathbb{E}\Big[\|\bar{\bm{w}}_{g}^{(k)}-\bar{\bm{w}}_{g}^{(k-1)}\|^2_2\Big]\notag\\
				& \le  -\frac{\eta T}{2} \mathbb{E}\|\nabla F(\bar{\bm{w}}_{g}^{(k-1)})\|_2^2 -\frac{\eta T^2}{2}(\frac{1}{2T}-L\eta)\mathbb{E}\|\nabla F(\bm{w}_g^{(k-1)})\|_2^2\notag\\
				& \quad + \frac{16T^3\eta^3L^2}{N^2}(\frac{3}{2}-LT\eta)\|\nabla \bm F(\textbf{W}_g^{(k-1)})\|_F^2\notag\\
				&\quad +\frac{1}{N}\big[4T^2\eta^3\chi^2L^2 +LT\eta^2\chi^2+4\eta^3 T^3\chi^2L^2(\frac{1}{2T}-L\eta)\big]\notag\\
				& \quad + {\eta T}\mathbb{E}\|\nabla F(\bar{\bm{w}}_{g}^{(k-1)})-\nabla F(\bm{w}_g^{(k-1)})\|_2^2 \notag\\
				&\overset{(a)}\le -\frac{\eta T}{2} \mathbb{E}\|\nabla F(\bar{\bm{w}}_{g}^{(k-1)})\|_2^2 -\frac{\eta T^2}{2}(\frac{1}{2T}-L\eta)\mathbb{E}\|\nabla F(\bm{w}_g^{(k-1)})\|_2^2 \notag \\
				& \quad + {\eta T}\mathbb{E}\|\nabla F(\bar{\bm{w}}_{g}^{(k-1)})-\nabla F(\bm{w}_g^{(k-1)})\|_2^2\notag \\
				& \quad +\frac{16T^3\eta^3L^2}{N}(\frac{3}{2}-LT\eta)\times (4\mathbb{E}\|\nabla F(\bm{w}_g^{(k-1)})\|_2^2+4\kappa^2) \notag \\
				&\quad +\frac{16T^3\eta^3L^2}{N}(\frac{3}{2}-LT\eta)\times\frac{8L^2}{N}\sum_{i\in\mathcal{V}}\mathbb{E}\|\bar{\bm{w}}_g^{(k-1)}-\bm{w}_{i,g}^{(k-1)}\|_2^2 \notag \\
				& \quad + \frac{1}{N}\big[4T^2\eta^3\chi^2L^2 +LT\eta^2\chi^2+4\eta^3 T^3\chi^2L^2(\frac{1}{2T}-L\eta)\big],
			\end{align}
			where (a) follows from Lemma \ref{l11}.
			The third term in the right side of the last inequality of (\ref{pd32_5}) can be further bounded by
			\begin{align}
				\label{pd32_6}
				&{\eta T}\mathbb{E}\Big\|\nabla F(\bar{\bm{w}}_g^{(k-1)})-\frac{1}{N}\sum_{i\in\mathcal{V}}\nabla f_i(\bm{w}_{i,g}^{(k-1)})\Big\|^2_2 \notag\\
				= & \frac{\eta T}{N^2}\mathbb{E}\Big\|\sum_{i\in\mathcal{V}}\big[f_i(\bar{\bm{w}}_g^{(k-1)})-\nabla f_i(\bm{w}_{i,g}^{(k-1)})\big] \Big\|^2_2 \notag \\
				\le  & \frac{\eta T}{N}\sum_{i\in\mathcal{V}}\mathbb{E}\|f_i(\bar{\bm{w}}_g^{(k-1)})-\nabla f_i(\bm{w}_{i,g}^{(k-1)})\|_2^2 \notag \\
				\le & \frac{\eta T L^2}{N}\sum_{i\in\mathcal{V}}\mathbb{E}\|\bar{\bm{w}}_g^{(k-1)}-\bm{w}_{i,g}^{(k-1)}\|_2^2.
			\end{align}
			Thus, rearrange the terms and we can give the upper bound as follows.
			\begin{align}
				\label{pd32_7}
				&\mathbb{E}\Big[\nabla F(\bar{\bm{w}}_{g}^{(k-1)})^T(\bar{\bm{w}}_{g}^{(k)}-\bar{\bm{w}}_{g}^{(k-1)})\Big] +\frac{L}{2}\mathbb{E}\Big[\|\bar{\bm{w}}_{g}^{(k)}-\bar{\bm{w}}_{g}^{(k-1)}\|^2_2\Big]\notag\\
				& \le -\frac{\eta T}{2} \mathbb{E}\|\nabla F(\bar{\bm{w}}_{g}^{(k-1)})\|_2^2 -\frac{\eta T}{2}(\frac{1}{6}-LT\eta)\mathbb{E}\|\nabla F(\bm{w}_g^{(k-1)})\|_2^2 \notag \\
				& \quad + \frac{4\eta TL^2}{3N}\sum_{i\in\mathcal{V}}\mathbb{E}\|\bar{\bm{w}}_g^{(k-1)}-\bm{w}_{i,g}^{(k-1)}\|_2^2\notag \\
				& \quad +\frac{1}{N}\big[\eta^2LT(4\kappa^2T+\chi^2) + 6T^2\eta^3\chi^2L^2\big],
			\end{align}
			where the inequality follows from $\eta<\frac{1}{24TL}$.
		\end{proof}
		
		\subsection*{D. Derivation of (33)}
		In this part, we present the detailed derivation of (33) based on the given Lemmas in [30]. 
		For the D-SGD updating rule of the global model parameters, the main difference of our method from that in [30] lies in the multiple steps of local SGD in each round. Since the proposed method and analysis in [30] is based on single step SGD, we need to deal with those Lemmas by deriving them with some techniques, to get the results for multiple steps of SGD. 
		To this end, we first give the following Lemma \ref{l9}, \ref{l10} and \ref{l12} corresponding to Lemma 2, 3 and 5 in [30]. Additionally, the above Lemma \ref{l11} corresponds to Lemma 4 in [30]. These lemmas are derived with adjustment and derivation for multiple steps of SGD.
		\begin{lemma}
			\label{l9}
			({Residual error decomposition}) Let $\bm{W}_{g}^{(0)} = \bm{w}_{g}^{(0)}\bm{1}_N^T$ be the initialization, i.e., the global model parameters for all nodes are initialized to the same values. We define $\bar{\bar{\bm{A}}}_{s, k-1} = \prod_{l=s}^{k-1}\tilde{\bm{A}}$ and $\bm{Q}=\frac{1}{N}\bm{1}_N\bm{1}_N^T$.
			If we apply the updating rule (10), then for any $k\ge 2$, we have 
			$$
			\bm{W}_{g}^{(k)}(\bm{I}_N-\bm{Q}) = -\eta\sum_{s=1}^{k-1}\Xi_{g}^{(s)}(\bar{\bar{\bm{A}}}_{s, k-1}-\bm{Q}).
			$$
		\end{lemma}
		Lemma \ref{l9} can be easily derived according to Lemma 2 in [30], where the one step gradient is replaced by the accumulated gradients with multiple steps.
		
		\begin{lemma}
			\label{l10}
			(Gradient variance decomposition) Given any sequence of deterministic matrices $\{\textbf{B}_s\}_{s=1}^k$, then for any $k\ge1$,
			\begin{align}
				& \mathbb{E}\Big\|\sum_{s=1}^k\big[\Xi_{g}^{(s)}-\sum_{t=0}^{T-1}\nabla \bm F(\textbf{W}_{g,t}^{(s)})\big]\textbf{B}_s\Big\|_F^2\notag \\
				& = \sum_{s=1}^k\mathbb{E}\Big\|\big[\Xi_{g}^{(s)}-\sum_{t=0}^{T-1}\nabla \bm F(\textbf{W}_{g,t}^{(s)})\big]\textbf{B}_s\Big\|_F^2.
			\end{align}
		\end{lemma}
		Lemma \ref{l10} can also be easily derived according to the proof of Lemma 3 in [30], where the one step gradient is also replaced by the accumulated gradient, which subtracts its unbiased estimation.
		
		\begin{lemma}
			\label{l12}
			(Bound on residual errors). Let $\rho_{s, k-1}=\|\bar{\bar{\textbf{A}}}_{s, k-1}-\textbf{Q}\|$, where $\bar{\bar{\textbf{A}}}$ is defined in Lemma \ref{l9}. Then the residual error can be upper bounded, i.e.,
			\begin{align}
				& \frac{1}{N}\sum_{i\in\mathcal{V}}\mathbb{E}\|\bm{w}_{i,g}^{(k)}-\bar{\bm{w}}_g^{(k)}\|_2^2\notag\\
				& \le 2\eta^2\sum_{s=1}^{k-1}\rho_{s,k-1}^2T\chi^2 \notag\\
				&\quad + 2\eta^2\Big(\sum_{s=1}^{k-1}\rho_{s, k-1}\Big)\Big(\sum_{s=1}^{k-1}\frac{3\rho_{s, k-1}T^2}{N}\mathbb{E}\|\nabla \bm F(\textbf{W}_{g}^{(s)})\|_F^2\Big) \notag\\
				&\quad +2\eta^2\Big(\sum_{s=1}^{k-1}\rho_{s, k-1}\Big)\Big(\sum_{s=1}^{k-1}8\rho_{s, k-1}T^3\eta^2\chi^2L^2\Big).
			\end{align}
		\end{lemma}
		The proof of Lemma \ref{l12} is a little complicated and here we give its proof as follows, with some adjustment based on the proof of Lemma 5 in [30].
		\begin{proof}
			\begin{align}
				& \sum_{i\in\mathcal{V}}\mathbb{E}\|\bm{w}_{i,g}^{(k)}-\bar{\bm{w}}_g^{(k)}\|_2^2 =\mathbb{E}\|\textbf{W}_{g}^{(k)}(\textbf{I}_N-\textbf{Q})\|_F^2 \notag \\
				& \overset{(a)}= \eta^2\mathbb{E}\Big\|\sum_{s=1}^{k-1}\Xi_{g}^{(s)}(\bar{\bar{\bm{A}}}_{s, k-1}-\bm{Q})\Big\|_F^2 \notag \\
				& \le 2\eta^2\mathbb{E}\Big\|\sum_{s=1}^{k-1}\big(\Xi_{g}^{(s)}-\sum_{t=0}^{T-1}\nabla \bm F(\textbf{W}_{g,t}^{(s)})\big)(\bar{\bar{\bm{A}}}_{s, k-1}-\bm{Q})\Big\|_F^2 \notag \\
				&\quad +2\eta^2\mathbb{E}\Big\|\sum_{s=1}^{k-1}\sum_{t=0}^{T-1}\nabla \bm F(\textbf{W}_{g,t}^{(s)})(\bar{\bar{\bm{A}}}_{s, k-1}-\bm{Q})\Big\|_F^2 \notag\\
				&\overset{(b)}= 2\eta^2\sum_{s=1}^{k-1}\mathbb{E}\Big\|\big(\Xi_{g}^{(s)}-\sum_{t=0}^{T-1}\nabla \bm F(\textbf{W}_{g,t}^{(s)})\big)(\bar{\bar{\bm{A}}}_{s, k-1}-\bm{Q})\Big\|_F^2 \notag\\
				&\quad + 2\eta^2\mathbb{E}\Big\|\sum_{s=1}^{k-1}\sum_{t=0}^{T-1}\nabla \bm F(\textbf{W}_{g,t}^{(s)})(\bar{\bar{\bm{A}}}_{s, k-1}-\bm{Q})\Big\|_F^2 \notag\\
				& \overset{(c)} \le2\eta^2\sum_{s=1}^{k-1}\mathbb{E}\Big\|\big(\Xi_{g}^{(s)}-\sum_{t=0}^{T-1}\nabla \bm F(\textbf{W}_{g,t}^{(s)})\big)(\bar{\bar{\bm{A}}}_{s, k-1}-\bm{Q})\Big\|_F^2 \notag\\
				&\quad + 2\eta^2\mathbb{E}\Big(\sum_{s=1}^{k-1}\|\sum_{t=0}^{T-1}\nabla \bm F(\textbf{W}_{g,t}^{(s)})(\bar{\bar{\bm{A}}}_{s, k-1}-\bm{Q})\|_F\Big)^2 \notag\\
				& \overset{(d)}\le 2\eta^2\mathbb{E}\sum_{s=1}^{k-1}\big\|\Xi_{g}^{(s)}-\sum_{t=0}^{T-1}\nabla \bm F(\textbf{W}_{g,t}^{(s)})\big\|_F^2\big\|\bar{\bar{\bm{A}}}_{s, k-1}-\bm{Q}\big\|_2^2\notag\\
				&\quad +2\eta^2\mathbb{E}\Big(\sum_{s=1}^{k-1}\big\|\sum_{t=0}^{T-1}\nabla \bm F(\textbf{W}_{g,t}^{(s)})\big\|_F\big\|(\bar{\bar{\bm{A}}}_{s, k-1}-\bm{Q})\big\|\Big)^2 \notag\\
				& \overset{(e)}= 2\eta^2\sum_{s=1}^{k-1}\rho_{s,k-1}^2\mathbb{E}\big\|\Xi_{g}^{(s)}-\sum_{t=0}^{T-1}\nabla \bm F(\textbf{W}_{g,t}^{(s)})\big\|_F^2 \notag\\
				&\quad + 2\eta^2\mathbb{E}\Big(\sum_{s=1}^{k-1}\rho_{s, k-1}\|\sum_{t=0}^{T-1}\nabla \bm F(\textbf{W}_{g,t}^{(s)})\|_F\Big)^2 \notag\\
				&\overset{(f)}\le 2\eta^2\sum_{s=1}^{k-1}\rho_{s,k-1}^2NT\chi^2 \notag\\
				&\quad + 2\eta^2\Big(\sum_{s=1}^{k-1}\rho_{s, k-1}\Big)\Big(\sum_{s=1}^{k-1}\rho_{s, k-1}\mathbb{E}\|\sum_{t=0}^{T-1}\nabla \bm F(\textbf{W}_{g,t}^{(s)})\|_F^2\Big),
			\end{align}
			where (a) follows from Lemma \ref{l9}; (b) follows from Lemma \ref{l10}; (c) follows from the triangle inequality $\|\sum_{s=1}^{k-1}\textbf{A}_s\|_F\le \sum_{s=1}^{k-1}\|\textbf{A}_s\|_F$; (d) follows from the inequality $\|\textbf{A}\textbf{B}\|_F\le \|\textbf{A}\|_F\|\textbf{B}\|$ for matrix $\textbf{A}$ and $\textbf{B}$; (e) follows from the notation $\rho_{s, k-1}=\|\bar{\bar{\textbf{A}}}_{s, k-1}-\textbf{Q}\|$; (f) follows from Assumption 4 and Cauchy inequality.
			Then according to Lemma \ref{l13}, we can further get
			\begin{align}
				& \sum_{i\in\mathcal{V}}\mathbb{E}\|\bm{w}_{i,g}^{(k)}-\bar{\bm{w}}_g^{(k)}\|_2^2 \notag \\
				&\le 2\eta^2\sum_{s=1}^{k-1}\rho_{s,k-1}^2NT\chi^2 \notag\\
				&\quad + 2\eta^2\Big(\sum_{s=1}^{k-1}\rho_{s, k-1}\Big)\Big(\sum_{s=1}^{k-1}3\rho_{s, k-1}T^2\mathbb{E}\|\nabla \bm F(\textbf{W}_{g}^{(s)})\|_F^2\Big) \notag\\
				&\quad +2\eta^2\Big(\sum_{s=1}^{k-1}\rho_{s, k-1}\Big)\Big(\sum_{s=1}^{k-1}8\rho_{s, k-1}T^3\eta^2\chi^2L^2N\Big).
			\end{align}
		\end{proof}
		
		It can be observed that compared with the upper bound in Lemma 5 in [30], in our result, the first term multiples $T$, the coefficient before $\mathbb{E}\|{\nabla F}(\textbf{W}_g^{(k-1)})\|_F^2$ multiples $3T^2$, and we additionally have one constant term. Then it is easy to derive the following results based on the proof of Lemma 6 in [30].
		\begin{align}
			&\frac{1}{N}\sum_{i\in\mathcal{V}}\sum_{k=1}^{K}\mathbb{E}\|\bar{\bm{w}}_g^{(k-1)}-\bm{w}_{i,g}^{(k-1)}\|_2^2\notag\\
			\le & \Big[A_K\chi^2+B_KT( \kappa^2+T\eta^2\chi^2L^2)+\notag\\
			&\quad \frac{C_KT}{K}\sum_{k=1}^{K}\mathbb{E}\big\|{\nabla F}(\bm{w}_{g}^{(k-1)})\big\|_2^2\Big]\times \frac{24\eta^2TK}{1-48\eta^2L^2T^2C_K},
		\end{align}
		where $A_K, B_K, C_K, D_K$ are defined as follows.
		\begin{align}
			&A_K=\frac{1}{K}\sum_{k=1}^{K}\sum_{s=1}^{k-1}\rho_{s, k-1}^2,\quad B_K=\frac{1}{K}\sum_{k=1}^{K}\Big(\sum_{s=1}^{k-1}\rho_{s, k-1}\Big)^2,\notag \\
			& C_K = \max_{s\in[K-1]}\sum_{k=s+1}^{K}\rho_{s, k-1}\Big(\sum_{l=1}^{k-1}\rho_{l, k-1}\Big). \notag
		\end{align}

\end{document}